\documentclass[12pt]{aa}

\usepackage{graphicx,amssymb,amsmath,times}

\usepackage{longtable}

\usepackage{bm} 

\usepackage[draft]{hyperref}

\usepackage[varg]{txfonts}
\usepackage[T1]{fontenc}

\usepackage{natbib}

\bibpunct{(}{)}{;}{a}{}{,} % to follow the A&A style

\usepackage[normalem]{ulem}

\def\simlt{\lower.5ex\hbox{\ltsima}}
\def\simgt{\lower.5ex\hbox{\gtsima}}

\def\gtsim{\;\lower.6ex\hbox{$\sim$}\kern-6.7pt\raise.4ex\hbox{$>$}\;}
\def\ltsim{\;\lower.6ex\hbox{$\sim$}\kern-6.9pt\raise.4ex\hbox{$<$}\;}

\def\sec{${}^{\prime\prime}$}

\def\bmv{\hbox{\it B--V\/}}

\def\ngc#1{\hbox{NGC$\,$#1}}
\def\wcen{$\omega$ Cen}

\titlerunning{The RR Lyrae and Type II Cepheids of $\omega$ Cen}
\authorrunning{Braga et al.}
\usepackage{color}

\begin{document}

% SCRIVERLO A MANO SE CI SONO I THANKS!!!!!!!
\title{On the separation between RR Lyrae and Type II Cepheids 
and their importance for distance determination: 
the case of omega Cen\thanks{Tables 1, 2, 3, 7, A.1,
B.1 and B.2 are only available in electronic form at 
the CDS via anonymous ftp to \url{cdsarc.u-strasbg.fr} 
(130.79.128.5) or via \url{http://cdsweb.u-strasbg.fr/cgi-bin/qcat?J/A+A/xxx}} }

%%%%%%%%%%%%%%%%%%%%%%%%%%%%%%%%%%%%%%%%%%%%%%%%%%   
% generate_latex_authors,['*Braga*','*Bono*','*Fiorentino*','*Stetson*','*Dall*ora*','*Salaris*','*da Silva*','*Fabrizio*','*Marinoni*','*Marrese*','*Mateo*','*Matsun*','*Monelli*','*Wallerst*'],1 
%%%%%%%%%%%%%%%%%%% TITLE PAGE %%%%%%%%%%%%%%%%%%%   

% % ---Gruppo principale---
%    \author{V.F.~Braga\inst{1,2} 
%    \and X.~YZ\inst{2}}
% 
% \institute{INAF-Osservatorio Astronomico di Roma, via Frascati 33, 00040 Monte Porzio Catone, Italy\\
% \and Space Science Data Center, via del Politecnico snc, 00133 Roma, Italy
% }

\author{
V.~F.~Braga \inst{1,2}
\and G.~Bono \inst{1,3}
\and G.~Fiorentino \inst{1}
\and P.~B.~Stetson \inst{4}
\and M.~Dall'Ora \inst{5}
\and M.~Salaris\inst{6}
\and R.~da Silva\inst{1,2}
\and M.~Fabrizio\inst{1,2}
\and S.~Marinoni\inst{1,2}
\and P.~M.~Marrese\inst{1,2}
\and M.~Mateo \inst{7}
\and N.~Matsunaga \inst{8}
\and M.~Monelli \inst{9}
\and G.~Wallerstein\inst{10}
}
\institute{
INAF-Osservatorio Astronomico di Roma, via Frascati 33, 00040 Monte Porzio Catone, Italy\\
\and Space Science Data Center, via del Politecnico snc, 00133 Roma, Italy\\
\and Dipartimento di Fisica, Universit\`a di Roma Tor Vergata, via della Ricerca Scientifica 1, 00133 Roma, Italy\\
\and Herzberg Astronomy and Astrophysics, National Research Council, 5071 West Saanich Road, Victoria, British Columbia V9E 2E7, Canada\\
\and INAF-Osservatorio Astronomico di Capodimonte, Salita Moiariello 16, 80131 Napoli, Italy\\
\and Astrophysics Research Institute, Liverpool John Moores University, IC2, Liverpool Science Park, 146 Brownlow Hill, Liverpool,L3 5RF, UK\\
\and Department of Astronomy, University of Michigan, Ann Arbor, MI, USA\\
\and Department of Astronomy, The University of Tokyo, 7-3-1 Hongo, Bunkyo-ku, Tokyo 113-0033, Japan\\
\and Instituto de Astrof\'isica de Canarias, Calle Via Lactea s/n, E38205 La Laguna, Tenerife, Spain\\
\and Deptartment of Astronomy, University of Washington, Seattle, WA 98195, USA
}
\date{\centering Submitted \today\ / Received / Accepted }

\abstract{The separation between RR Lyrae (RRLs) and Type II 
Cepheid (T2Cs) variables based on their period is debated. Both 
types of variable stars are distance indicators and
we aim to promote the use of T2Cs as distance indicators 
in synergy with RRLs. We adopted new and existing 
optical and Near-Infrared (NIR) photometry 
of \wcen~to investigate several diagnostics 
(colour-magnitude diagram, Bailey diagram, 
Fourier decomposition of the light curve, amplitude ratios) for
their empirical separation. We found that the classical period threshold at 1 day
is not universal and does not dictate the evolutionary stage: V92 
has a period of 1.3 days but is likely to be still in its core 
Helium-burning phase, typical of RRLs. We also derived NIR
Period-Luminosity relations and found a distance modulus 
of 13.65$\pm$0.07 (err.)$\pm$0.01 ($\sigma$) mag, in agreement
with the recent literature. We also found that RRLs and T2Cs 
obey the same PL relations in the NIR. This equivalence gives the opportunity
to adopt RRLs+T2Cs as an alternative to classical Cepheids
to calibrate the extragalactic distance scale.}

\maketitle

%_______________________________________________________________________________
\section{Introduction} \label{chapt_intro_omega}
%_______________________________________________________________________________

Type II Cepheids (T2Cs) are pulsating variable stars of the 
Cepheid Instability Strip (IS),  
typically associated to old ($>$10 Gyr) stellar populations. 
They are low-mass stars in either the 
post-horizontal branch (post-HB), asymptotic giant branch (AGB) or 
post-AGB phase \citep[see, e.g.,][Bono et al., 2020, submitted]{gingold74,sweigart89,bono97e}
and are mostly found in stellar systems with extended 
Blue HBs (BHB). T2Cs can be used as standard candles for distance 
estimates because they display optical and near-infrared (NIR) 
Period-Luminosity (PL) relations, analogous to those of RR Lyrae (RRLs) and 
Classical Cepheids (CCs). Moreover, PL relations 
of T2Cs are only minimally affected by metal abundance 
\citep{bono97e,matsunaga06,dicriscienzo07,matsunaga2013,lemasle15}.

Despite being much less numerous than RRLs \citep[the 
Galactic Bulge hosts almost 70,000 RRLs but only around 
1,000 T2Cs][]{soszynski14,soszynski2017,soszynski2019}, T2Cs 
are from one to five mag brighter. 
This means that their detection in high extinction environments
\citep[e.g., the Galactic Bulge,][]{bhardwaj17c,braga2018b,braga2019b}
is easier, but also that they can be identified and characterized in external 
galaxies. Indeed, T2Cs have been found near M31 \citep{kodric2018} in  M101 
and M106 \citep{stetson98a,macri06,majaess2009}, and more recently even 
an RVT in the Seyfert 1 galaxy NGC~4151 \citep{yuan2020}.

In particular, in the NIR bands ($JHK_s$) T2Cs seem 
to follow the same PL relations of RRLs \citep{majaess2010}
although there is still no solid empirical evidence.
This means that RRLs and T2Cs may potentially be adopted jointly to 
calibrate Supernovae Type Ia (SNIa) luminosities and, in turn, to 
measure the Hubble constant ($H_0$).
In the last years, a $\sim$4$\sigma$ level tension between estimates of the local value 
$H_0$ from the CCs+SNIa scale \citep[$H_0$=74.03$\pm$1.42 km/s/Mpc][]{riess2019}, and from  
CMB analyses \citep[$H_0$=67.39$\pm$0.54 km/s/Mpc][]{planck2018} has arisen,
meaning that either a bias in any of these techniques, 
or new physics come into play.
Note, that the discussion concerning the extent of the current tension is far 
from being settled. In a recent investigation \citet{majaess2020} argued that 
neglected or inaccurate blending corrections may result in an overestimated $H_0$.
Moreover, by using a new calibration of SNIa based on the luminosity of the Tip of the Red Giant 
Branch \citep[TRGB,][]{beaton2019}, alleviates the tension by yielding
intermediate values of $H_0$, \citep[69.6$\pm$1.7 km/s/Mpc][]{freedman2020}.

An independent estimate
of the local $H_0$ obtained with RRLs+T2Cs could be crucial to 
either validate or reconsider the tension. To adopt RRLs+T2Cs
instead of CCs would also remove a population bias, for the latter variables
are only found in late type galaxies, while RRLs and T2Cs are
ubiquitous. A calibration of SNIa distances through old population 
tracers---RRLs and TRGB---has been already proposed \citep{beaton2016},  
however, this approach requires one more 
intermediate calibrations (that from RRLs to the TRGB),
brings into play different physics (TRGB stars are 
not pulsating variables) and, in turn, different systematics 
when compared with homogeneous PL relations for RRLs and T2Cs.

The separation between RRLs and 
T2Cs is a long-standing problem.  As 
a first approximation, it is possible to adopt a period 
threshold, whose exact value is still a matter of debate.
A threshold of $\sim$0.8 days was set in the review by
\citep{gautschysaio1996} where 
type 1 (AHB1) stars above the HB --as defined in 
\citet{strom1970} and \citet{diethelm1983,diethelm1990}-- 
were considered as T2Cs rather than evolved RRLs. 
This threshold is however obsolete, because a more
extended and homogeneous investigation, based on 
period distribution and on the Fourier 
parameters of the light curve of RRLs in the Galactic
Bulge, has now set the threshold at 1 day \citep{soszynski08c,soszynski14}.

RRLs in the Bulge have a primordial 
(or minimally enhanced) helium abundance \citep{marconiminniti2018} 
but there is theoretical evidence
that helium enhancement increases the periods of RRLs
\citep{marconi2018}. This means that a 1-day period threshold  
should be considered a particular case of a more generic
chemical- physics- and evolution-dependent threshold.

T2Cs are typically separated into BL Herculis (BLHs), 
W Virginis (WVs) and RV Tauri (RVTs) stars. 
\citet{soszynski2011} investigation with OGLE-III data  
found two minima---at 5 and 20 days---in
the period distribution of 335 T2Cs in the 
Galactic Bulge. They adopted these values
as thresholds between BLHs-WVs and WVs-RVTs, respectively. 
These thresholds were later validated 
on the OGLE-IV sample of bulge T2Cs, that is almost
three times larger \citep{soszynski2017}. 
Note that the BLHs-WVs threshold of the General Catalog 
of Variable Stars \citep[GCVS,][]{samus2017}, is 4 days, 
based on the period distribution of T2Cs in the LMC 
\citep{soszynski08c}. However, this is based on a 
small sample ($\sim$200 T2Cs) and new LMC data 
\citep[$>$300 T2Cs,][]{soszynski2018} invalidate  
the 4-day threshold.

% Based on a
% recent spectroscopic investigation \citet{kovtyukh2018b}
% propsed to further separate short-period BLHs 
% ($<$3 days) into two classes: metal-poor, 
% $\alpha$-enhanced ones, called UY Eridani (UYE)
% and metal-rich, Na-enhanced ones, ambiguously called BL Herculis.
% However, we will ignore this classification for two reasons: $i)--$, 
% we do not have spectroscopic abundances for the 
% short-period T2Cs in \wcen; $ii)--$ by visually inspecting 
% the light curves of the variables in
% \citep{kovtyukh2018b}, and comparing with OGLE atlas of 
% light curves, several of the former might be misclassified 
% and might belong to different Cepheid classes (CCs, Anomalous).

Whether RVTs should be all classified as bona-fide T2Cs 
is still a pending issue. In fact RVTs 
are associated to either low- or intermediate-mass 
\citep[from $\sim$0.5 up to $\sim$3$M_{\odot}$][]{dawson1979} 
post-AGB stars \citep{gingold1985,wallerstein2002}, belonging 
to old- and intermediate-age populations, respectively. 
To further stress the importance of the difference
between old and intermediate-age RVTs, 
a different naming was proposed for the low-mass ones 
\citep[][V2342 Sgr stars]{catelan15}. There is empirical 
evidence that RVTs in Galactic globular clusters (GGCs) which should belong to the 
V2342 Sgr class, do not share the same
properties as field RVTs \citep[][e.g., they are missing the typical 
alternating deep and shallow minima]{zsoldos1998}.
Finally, there is no consensus on the use of 
RVTs as reliable distance indicators.
In fact, it is still a matter of debate whether they do follow the PL
relation of BLHs and WVs \citep{matsunaga06,ripepi2015,bhardwaj17b}.
% \citet{ripepi2015} found that, in the NIR bands, RVTs are
% overluminous respect to the PL relation of BLHs and WVs. 
% The same overluminosity is less evident in the 
% optical bands \citep[][, SMC and LMC]{soszynski2018}. On the other hand, 
% \citet{matsunaga06} and \citet{bhardwaj17b} found that 
% RVTs do follow the same NIR ($JHK_s$) PL 
% relations as BLHs and WVs in GGCs and in the LMC. 
% The scenario is puzzling and provides 
% further evidence of the fact that RVTs in different 
% environments do have different properties and might 
% be ascribed to different populations.

Among nearby coeval stellar systems, the GGC \wcen~(NGC5139) is 
the best workbench for T2Cs.
It hosts the largest T2C sample in GGCs (seven T2Cs)
after the two more metal rich clusters
NGC6388 (twelve) and NGC6441 (eight), as well as long-period 
($>$0.7 days) RRLs. Moreover, three of its T2Cs have 
periods shorter than two days --which is optimal 
to investigate the transition between RRLs and T2Cs-- while 
all T2Cs in NGC6388 and NGC6441 have periods
longer than two days, with only one exception. 
While it is true that the Bulge hosts more RRLs and T2Cs, they are not at the 
same distance, the differential reddening and the stellar crowding are more 
severe, and NIR time series are only available for the $K_s$-band.

Furthermore, $\omega$ Cen is characterized by a well known spread in 
metallicity \citep{johnson_e_pilachowski2010,johnson2020}, in 
helium content \citep{lee1999,calamida2020}
and affected by peculiar radial distribution of metal-poor and 
metal-rich stellar populations 
\citep[][and references therein]{lee1999,calamida2020}.
This is an advantage because the pulsation properties 
of RRLs and T2Cs depend on chemical composition, meaning
that \wcen~is the GGC where these variable stars
have more heterogeneous and varied pulsation properties.

Recently, the unprecedented 
wealth of kinematic data from Gaia DR2 and 
APOGEE DR14  \citep{gaia_dr2,apogee_dr14} 
were employed by \citet{ibata2019} to validate the existence 
of the tidal stellar stream of \wcen. 
Based on its motion, \citet{myeong2019}  
argued that \wcen~could have been accreted
by the Milky Way during the Sequoia merger, 
while \citet{massari2019} and \citet{kruijssen2020}, more conservatively,  
associated \wcen~to either Sequoia or to 
the larger Gaia-Enceladus \citep{helmi2018} merger event.
On the theoretical side, \citet{bekki2019} validated these 
hypotheses and speculated that \wcen~was a GGC of an
accreted galaxy. Based on the properties of the different
stellar population of \wcen, \citet{calamida2020} proposed
that, before being accreted by the Milky Way, \wcen~might 
have formed by mergers between clusters and --eventually-- the 
nucleus of a dwarf galaxy.

Therefore, \wcen~is not only a GGC showing optimal properties
to investigate T2Cs, but also 
a very interesting object on its own, being the largest 
and most heterogeneous GGC within the Galaxy.
Finally, being close and well populated,  
its distance was estimated by using several 
diagnostics (RRLs, T2Cs, TRGB, white dwarfs, eclipsing binaries).

% The last multi-band photometric study of T2Cs 
% in \wcen~dates back to \citet{gonzalez1994b} but did not 
% include the two variables V60 and V61. 
% and the only spectroscopic
% study of these objects was performed in the same year
% \citep{gonzalez1994b}, yielding abundance estimates for more than 
% 20 elements in V1, V29 and V48, which are the three 
% \wcen~T2Cs with the longest periods (larger than 4 days).

The aim of the paper is to provide more rigorous criteria
to differentiate between RRLs and T2C, by viewing them from
an evolutionary perspective. This would be a complete 
reversal of the point of view. In fact, until now, the 
period threshold has been the most common criterion to 
separate RRLs and T2Cs. Almost always,
this separation separates HB stars on one side (P$<$1 day) and
post-HB stars on the other side (P$>$1 day). As a consequence,
RRLs are considered as HB (core He burning) stars 
and T2Cs are considered as post-HB (double shell burning) stars.
Our claim is, instead, that the leading argument to separate
RRLs and T2Cs is their evolutionary stage, and that the 
empirical separation---only based on pulsation 
properties---is the consequence.

Moreover, we plan to adopt T2Cs as distance indicators, compare
them with RRLs and use them jointly within a common distance diagnostic.
This means that, on the one side, we aim for
a more solid understanding of the differences in the evolutionary
properties of both RRLs and T2Cs and, on the other side,
we use them together as a single distance indicator.
Seemingly, the two aims are at odds, but this is not the case. In
fact, for distance determinations (especially of Local Group
galaxies), it would be a threefold advantage to
adopt a common PL relation because $i)$ RRLs complement
the small number of T2Cs; $ii)$ T2Cs complement the
lower brightness of RRLs, and $iii)$ together they provide a
wider period range on which to calibrate the relation upon, as 
it was already suggested by \citet{benedict11}.
Still, RRLs and T2Cs are not the same objects from the evolutionary
point of view, and it is important to provide a clear criterion to
tell them apart not only for a mere taxonomical purpose, but also
for the correct development of pulsation models (Bono et al.
2020, submitted) and to investigate the population 
ratios of the host stellar system (GCs, nearby galaxies...).

The paper is structured as follows: in Section~\ref{chapt_t2c} we 
discuss the pulsation properties of T2Cs and long-period RRLs
in \wcen; moreover, we discuss the transition between the two
types of variables, also based on Bulge T2Cs. 
Section~\ref{chapt_comparison} is devoted to the 
comparison of the Bailey diagram and amplitude ratios of 
the T2Cs in \wcen, with those in the Galactic Halo and Bulge.
We discuss the PL relations of T2Cs,
their comparison with the PLs of RRLs, 
and use them to estimate the distance of \wcen~in 
Section~\ref{chapt_distance_omega}. We discuss 
our results in Section~\ref{section_conclusion}.

% 1) Confronto con tutti gli altri ammassi 
% globulari della regione P > 0.85
% Importante non inserire RRL da GGC che non
% hanno T2C perchè partiamo da HB diversi XXX

%_______________________________________________________________________________
\section{Type II Cepheids}\label{chapt_t2c}

According to the catalogue of variable stars in GGCs 
by \citet{Clement01}, \wcen~is the third richest GGC in T2Cs, after
the metal-rich clusters NGC~6388 and NGC~6441. In fact, \wcen~
hosts seven T2Cs: five BLHs (V43, V48, V60, V61, V92), one WV 
(V29) and one RVT (V1). All these variables were already
discovered in the first investigation of variable
stars in \wcen~\citep{bailey1902}.

A detailed list of the T2Cs and their pulsation properties
is provided in Appendices \ref{chapt_rrind_omega} and \ref{par:omegacen_lightcurves}.

% \subsection{Subclasses of T2Cs}\label{t2c_definition}

% {\it RVT generally show large IR colour excess 
% (Gehrz 1972). Dust envelope? (Wallerstein 2002)
% XXX IR EXCESS DI V1 con WISE: circa 0...
%_______________________________________________________________________________

\subsection{The transition between RRLs and T2Cs} \label{section_rrlt2c}

The period threshold between RRLs and T2Cs is a long-standing
dilemma in the field of pulsating variables. In the 
literature, the accepted values for the maximum period of RRLs range from 
0.75 days \citep{wallersteincox1984} to 2.5 days \citep{diethelm1983}. 

An investigation of a homogeneous and extended sample
of RRLs and T2Cs in the LMC with $VI$-band OGLE photometry, 
allowed to establish a more solid empirical threshold at 
1 day \citep{soszynski08c}. This was based on the period 
distribution and on the position 
of the pulsating variables in the $\log{P}$-$\phi_{21}$ plane, 
where $\phi_{21}$ is one of the coefficients of the Fourier 
series fit to the light-curve.

As stated in the Introduction, the threshold at 1 day should be
considered as a lower limit in the case of a cosmological helium
abundance. In fact, RRL pulsation models \citep{marconi2018} predict 
longer pulsation periods for RRLs with helium-enhanced chemical composition, hence 
in stellar systems hosting 
helium-enhanced stellar populations, the 
1-day threshold might not be reliable, and the separation between 
RRLs and BLHs might be less sharp. 
Recently, \citet{kovtyukh2018a} hypothesised that
the very existence of BLHs is to be ascribed to helium 
enhancement in a progenitor mass of 0.8 M$_{\odot}$. 
This is consistent with the fact that the
BHB of the metal-rich GGCs NGC6441 and NGC6388 
(the two GGCs hosting the highest number of T2Cs) can be reproduced by
helium-enhanced (Y=0.35-0.40) HB models\citep{busso2007,bellini2013}.
\citet{dalessandro11b} and \citet{tailo2019} 
showed that the initial helium content also affects
the mass loss on the RGB and, in turn, the position of the stars
on the HB. The more helium-enhanced 
is the progenitor, the least massive and bluer is the star 
on the ZAHB, both because the mass loss is higher
and because, at fixed age and fixed mass-loss rate, 
the turn-off mass is smaller.
Therefore, helium enhancement would favour an
evolution to T2Cs, which are observed only in systems  
with a well-populated BHB.
However, there is no evidence of an extensive 
helium-enhanced population in the Halo, thick disk and 
old population of the Galactic Bulge, where 
Galactic T2Cs are found. 
This means that, while helium enhancement 
might favour  the formation 
of T2Cs, it is not the only requirement. 
Moreover, evolved helium-enhanced RRLs 
could have periods and luminosities similar
to those of BLHs \citep{marconi2018}.

To this purpose, 
\wcen~is the most appropriate stellar system to inspect the 
RRL-BLH separation for several reasons.

$i)$ \wcen~displays a well-defined BHB \citep{castellani2007},
and T2Cs are associated with stellar systems showing an extended HB, 
since their progenitors are mainly BHB stars 
(\citealt{beaton2018}; Bono et al. 2020, submitted).
% This means that their formation is affected by the same physical 
% mechanisms impacting on the distribution of stars on the HB 
% \citep{beaton2018,kunder2018_issi,torelli2019}}

$ii)$,
It hosts five RRLs with periods between 0.85 and 1 day 
\citep{navarrete15,braga16} and four BLHs
with a period shorter than 3 days (V43, V60, V61 and V92);

$iii)$, there is evidence 
that the HB of \wcen~should be, at 
least in part, enhanced in Helium content 
\citep[][Y $\sim$ 0.28-0.38]{cassisi2009,bellini2013,tailo2016,latour2018}.

We investigate the transition
between RRLs and T2Cs by using the CMD, Bailey diagram and the Fourier
parameters of the light curve fit. Note that the latter two diagnostics are not 
used for the long-period RRLs, because our sampling of the 
time series is not optimal for variables with periods too close to 
1 day (V263 and NV366) and we have too few $I$-band phase 
points for the other variables.

Note that the rate of period change is not among the quoted diagnostics, because 
the sampling of our data does not allow us to provide accurate measures of this 
parameter. Recent theoretical results about the evolutionary channels 
producing T2Cs \citep[][]{bono2020b} show  
that a significant fraction of T2Cs evolve from the blue to the red
side of the HRD. These are
low-mass ($0.495 \leq M/M_{\odot} < 0.55$) horizontal branch stars that after the
central helium exhaustion evolve towards the AGB. In subsequent
evolutionary phases they move back to the blue 
towards the WD sequence, but this happens at higher 
luminosities (WVs, RVTs). These objects are characterized in their fainter limit by positive 
period derivatives. However, these models might also perform several gravo-nuclear
loops \citep{bono97b,bono97d,constantino16} in the HRD, either during the AGB phase
and/or in their approach to the WD cooling sequence. Some of these loops take
place inside the instability strip, and in turn, the period derivative can attain
both positive and negative values.

Evolutionary models also show that long-period RRLs evolve from the blue to 
the red (see Fig.~5 in Bono et al. 2020). This means that positive period 
derivatives do not help in discriminating between long-period RRLs and short 
period T2Cs. On the other hand, both positive and negative 
period derivatives can be univocally associated to T2Cs. The current empirical 
evidence indicates that period derivatives for the short-period BLHs of \wcen, 
are positive \citep{jurcsik2001}. Therefore, they cannot help us in 
separating BLHs from RRLs.

% Note that the PL relation is not a good diagnostic to discriminate
% between T2Cs and RRLs, since there is evidence in the 
% literature \citep{matsunaga06} that, within the uncertainties, 
% T2Cs follow the same NIR PL as RRab stars. This holds
% also for V1 in \wcen~(see Section~\ref{xxx}).

\subsubsection{Long-period RRLs}\label{section_longrrl}

We inspect five RRLs with P$>$0.85 days (V91, V104, V150, 
V263 and NV366). We do not include NV455 
because we do not have data for this star, 
that is located more than 40 arcmin from the center.
There is a clear period gap at $\sim$0.935-0.995 days, where 
no RRLs are found. Unfortunately, 
there are no strong arguments to assess whether the gap is 
real or just due to the poor 
statistics of long-period RRLs.

V91, V104 and V150)---These variables are 
all below the period gap, and their position in the 
$V$,$B-I$ and $K$,$B-K$ CMDs (see Fig.\ref{fig_cmd})
is consistent with being bona-fide RRLs. 
In the Bailey diagram, they are on the 
locus derived by performing
a linear fit of $Amp(V)$ vs period of the RRab stars of NGC6388
and NGC6441 ($Amp(V)=2.30-2.04\cdot P$, see Fig.\ref{fig:bailey}).

These clusters have been defined as Oosterhoff III in the 
literature, however, we decided to call them 
Oosterhoff 0 \citep[Oo0,][]{braga16}, because 
these GGCs are very metal-rich 
([Fe/H]$<$1 dex) and the progression in metallicity is
replicated by the numbers (Oo0=metal-rich; 
OoI=metal-poor; OoII=very metal-poor).

V263)---This star is above the period gap 
(P=$\sim$1.01 days), but its pulsation amplitude is
very small, and seem to agree well with the decreasing trend of 
RRab amplitudes at long periods (see Fig.~\ref{fig:bailey}). 
Despite its very long period, this star is consistent with
being an RRab, due to its position in the CMDs, well within
the magnitude range of HB stars with the same colour.

% !!!!!! IMPORTANTE !!!!!
% aggiungo alle curve di luce di NV366 dei punti fake per fare un fit migliore
% NV366_kal04_1.fas
% 0               0.9         14.4         0.      fake:
% 0               0.9         14.4         0.      fake:
% 0               0.8         14.28         0.      fake:
% 0               0.8         14.28         0.      fake:
% NV366_kal04_2.fas
% 0               0.8         14.85         0.      fake:
% 0               0.8         14.85        0.      fake:
% 0               0.99         15.05         0.      fake:
% 0               0.99         15.05         0.      fake:

NV366)---The light curve is heavily aliased because 
its period is 0.9999 days (above the period gap). 
In the optical bands the phases around minimum 
are missing. In the NIR bands, 
only the upper part of the decreasing branch is sampled. 
Therefore, both the optical and the NIR mean magnitudes 
which we have derived are probably underestimated, and 
the amplitudes are not reliable. However, 
in both CMDs, it is placed within the HB. Despite
being located $\sim$3.05 arcmin from the cluster center, in a very 
crowded region, a visual inspection of the
images did not outline any sign of blending. 
However, the crowding probably
affects the photometric calibration between our 
different optical datasets \citep{braga16}.

We note from the Bailey diagrams in Fig.\ref{fig:bailey},
that a sizeable sample of \wcen~RRLs, including the 
five ones discussed above, is placed on the locus of RRab 
stars in Oo0 clusters. 
The debate regarding the relation between the high
metallicity of Oo0 clusters and the properties 
of their HB and RRLs is still
unsettled. \citet{pritzl2002a} invokes a bimodal distribution of
metallicities in the Oo0 clusters, with RRLs and BHB stars 
belonging to the more metal-poor population.
In this scenario, HB stars evolve to the 
red, thus crossing the instability strip as metal-poor off-ZAHB RRLs. 
However, low-resolution spectra of RRLs in NGC 6441, provide 
high metallicities for these stars \citep{clementini05b}, 
thus invalidating the scenario by \citet{pritzl2002a}.

A similar scenario is described by \citet{tailo2016}
for \wcen: Based on population synthesis, they 
found that RRLs in \wcen~should mostly be 
metal-poor and evolved (off-ZAHB), with a smaller fraction of
metal-rich and fainter ZAHB RRLs; neither of these populations
is significantly enhanced in helium (Y$\leq$0.28).
% However, their evolutionary tracks \citep{dantona02} are quite dissimilar 
% to the BaSTI tracks by \citet{pietrinferni06a,pietrinferni06b}.}

\begin{figure}[htbp]
\centering
\includegraphics[width=9cm]{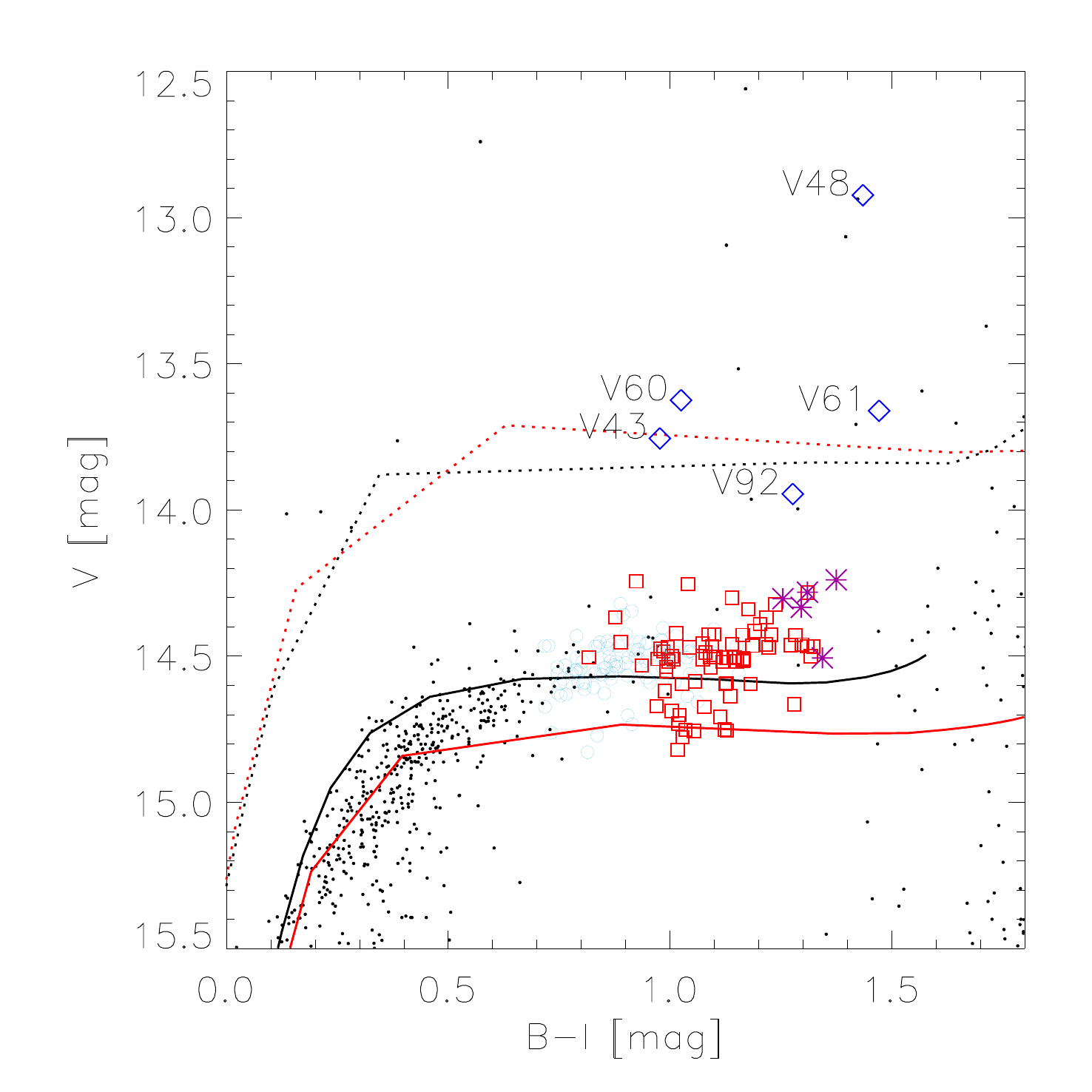}
\caption{Optical ($V$ vs $B-I$) CMD of \wcen. 
Light blue circles: RRc; red squares: RRab;
purple asterisks: long-period-RRab;
blue diamonds: BLHs. Black and red 
lines display the ZAHB (solid) and 
Helium-exhaustion (dotted) sequences for $\alpha$-enhanced, 
helium-normal HB models \citep{pietrinferni06b}
for two different metal 
contents: Z=0.0006 ([Fe/H]=--1.84) and 
Z=0.004 ([Fe/H]=--1.01).}
\label{fig:basti}
\end{figure}

We have compared the empirical distribution of the RRLs and
of the BHB in \wcen~with the BaSTI
$\alpha$-enhanced, helium-standard, 
Z=0.0006 and Z=0.004 tracks (corresponding 
to [Fe/H]$\sim$--1.84 and [Fe/H]$\sim$--1.01, 
respectively, see Fig.~\ref{fig:basti}).
The two metallicity values were chosen to 
reproduce the peak of the metallicity distribution
of RRLs \citep{magurno2019} and its metal-rich 
extension, which is the most prominent tail of the 
distribution. The majority of RRLs overlaps with 
the most metal-poor ZAHB model. 
On the other hand, the metal-rich ZAHB better 
fits the fainter and---presumably---more metal-rich RRLs.

% figura generata con temp_200217.pro 
% cp /home/vittorio/Documenti/Science/Basti_tracks/Tracks/hbz64aes_c04ae/ZAHBz64-eps-converted-to.pdf .
% /home/vittorio/Documenti/Science/Basti_tracks/Tracks/*/ZAHB*eps

We have checked that none of the Oo0-like RRLs in 
\wcen~ belong to the faint part of the RRL sample. %figura ooiii_rrls.pdf
Based on the analysis of high-resolution spectra, 
\citet{magurno2019} found indeed a minority of 
metal-poor RRLs. However, the 
latter are not significantly fainter, in the $V$ band, compared to the 
metal-poor RRLs, therefore, no firm conclusion can be reached.
%figura mpoor_vs_mrich.pdf

Therefore, while it is reasonable to assume that evolution 
is the most important factor in generating Oo0-like 
RRLs, secondary factors should be also considered to reproduce in detail
the observations. Also, their metallicity 
is not necessarily low.

One factor might be the initial helium abundance.
\citet{marconi11} investigated the period distribution of 
the RRLs in \wcen~and its correlation with helium enhancement; 
They concluded that the latter might be a concurrent reason 
for the long periods of RRLs in \wcen, but also put an upper limit
(20\%) to the fraction of helium-enhanced (Y$\geq$0.30) RRLs. 

Another factor might be the mass 
on ZAHB: since mass-loss along the RGB evolution is likely --at least partially--
a stochastic process, its correlation with Y or Z is not one-to-one.
Therefore, stars with the same chemical abundance, might lose
less/more than an average amount of mass on the RGB  
and will have cooler/higher temperatures on the ZAHB, 
as discussed by \citep{origlia2007,vanloon2008}.

\subsubsection{Short-period BLHs}\label{section_shortt2c}
% Per Diethelm, le CW sono diverse dalle BLH...
V43, V60, V61 and V92 are four BLHs with period 
shorter than 3 days. We have inspected their photometric 
properties to assess whether they might be better 
classified as very long-period RRLs, either 
evolved HB stars or helium-enhanced.
To this purpose, we adopt two different diagnostics, namely 
{\it evolutionary and pulsation
models} and {\it Fourier coefficients}.

{\it Evolutionary and pulsation models}: 
Figure~\ref{fig:basti} shows that V92 is consistently fainter than
the helium-exhaustion track. This is a further 
argument to reconsider its classification. We also 
point out that V43 is below the metal-rich 
helium-exhaustion track. However, we should assume a very high 
metallicity for V43 ([Fe/H]$>$1.3 dex, typical of 
less than 10\% of the RRL population of \wcen,
\citealt{magurno2019}) to consider its 
reclassification as candidate RRab, while none
of the other diagnostics points to a reclassification
as a borderline RRab/BLH star.
% Nonetheless, V43 is clearly of the AHB morphological class, 
% typically associated to RRab stars.

{\it Fourier coefficients}: We have 
derived the $\phi_{21}$ and $\phi_{31}$
Fourier coefficients of their $V$- and 
$I$-band light curves, to compare 
with those from the OGLE survey, which is the largest sample
of Fourier coefficients of pulsating variable stars.
OGLE provides only the $I$-band Fourier coefficients, hence 
by using OGLE $V$-band 
time series, we derived ourselves the Fourier coefficients of 
the $V$-band light curves of Bulge RRLs and T2Cs.

% Fourier Vband... possono servire per altre conclusioni
% visto che sono il primo che le fa?? Non a molto, visto 
% anche che non tutte le T2C hanno una buona copertura in V

\begin{figure}[htbp]
\centering
% \figurenum{1}
\includegraphics[width=8cm]{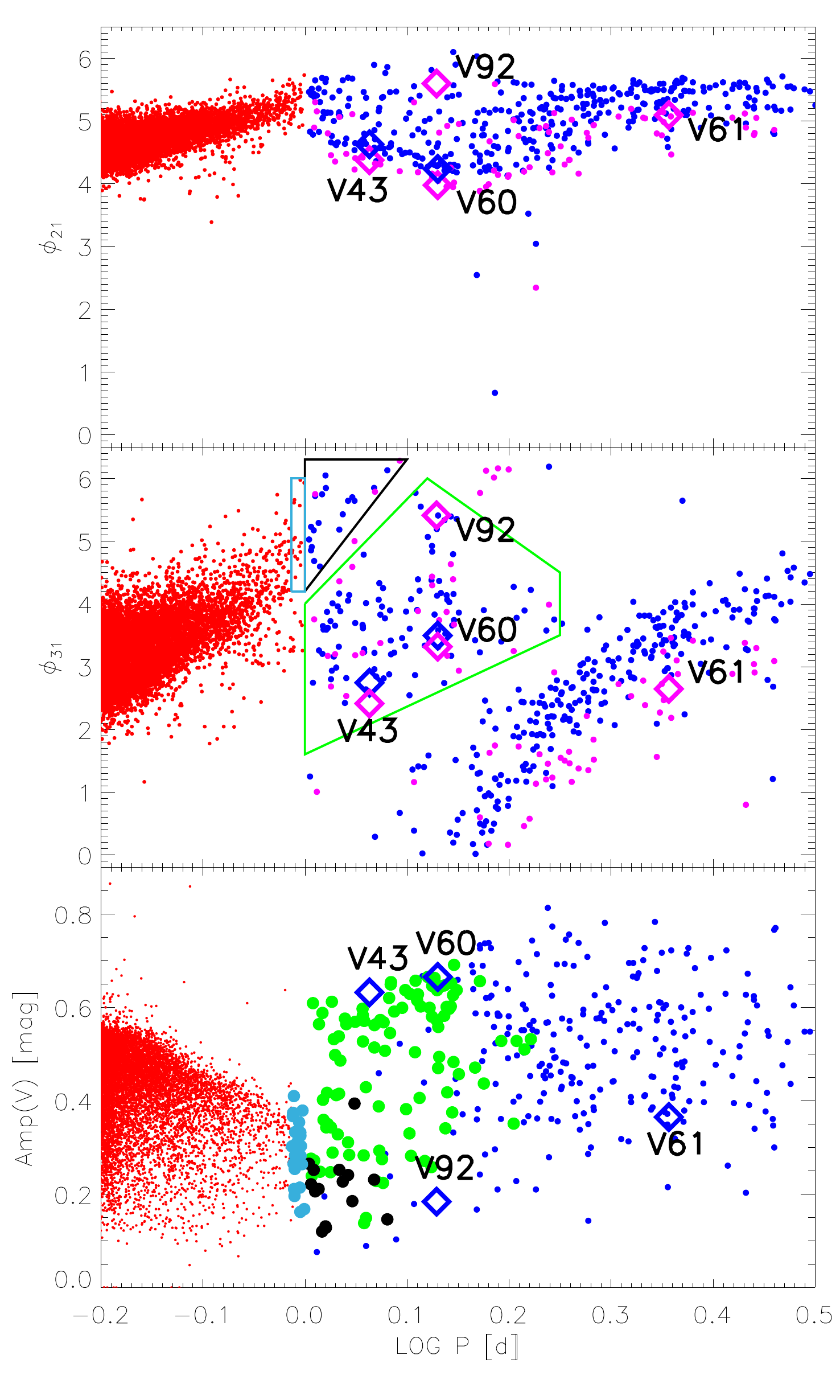} 
\caption{Top: $\phi_{21}$-$\log{P}$ diagram of 
long-period RRLs and short-period T2Cs in \wcen~and Galactic 
Bulge. Red dots: RRab; blue dots: T2Cs ($I$-band
Fourier coefficients); magenta dots: T2Cs ($V$-band
Fourier coefficients).
Middle: same as top but for the $\phi_{31}$ 
coefficients. The green box contains variables in the 
{\it short-period T2Cs} sample; the black box contains 
variables in the {\it candidate RRLs} sample, and the 
light blue box contains variables in the 
{\it RRL template} sample.
Bottom: Bailey diagram of the same variables as in 
the top and middle panel. Light blue, black and green 
cicles display variables in the {\it short-period T2Cs},
{\it candidate RRLs} and {\it RRL template} samples, 
respectively.}
\label{fig:phi}
\end{figure}

The top and middle panels of 
Fig.~\ref{fig:phi} display the $V$- (magenta)
and $I$-band (blue) Fourier coefficients of 
OGLE bulge RRLs and T2Cs in the $\phi_{21}$ and 
$\phi_{31}$ vs $\log{P}$ diagrams.

Larger symbols display the 
Fourier coefficients of the short-period BLHs in \wcen. 
Note that we did not derive the Fourier 
parameters of V61 and V92 in the $I$ band, given that their light
curve is not well sampled. However, we assume that their 
$I$-band coefficients are similar---or slightly higher---than their
$V$-band coefficients, as for all the other 
\wcen~and Bulge variables in the two planes.

%figura generata con IDLWorkspace/Default/Temp/temp_190316.pro

V43, V48 and V61 are well within
the typical loci of BLHs with the same period
within the $\phi_{21}$-$\log{P}$ and $\phi_{31}$-$\log{P}$ diagrams.
On the other hand, V92, especially in the $\phi_{21}$
vs $\log{P}$ plane, is at the edge of the Bulge BLH locus. Moreover,
its position is consistent with an extrapolation of the 
RRLs at longer periods. This triggers a question:
Could there be RRLs with periods longer than 1 day within the 
sample of T2Cs in the Bulge? To check this hypothesis, 
we have selected a few sub-samples of Bulge RRLs and T2Cs.

First, we selected 16 long-period (P>0.97 d) RRLs (light blue
box in the middle panel of Fig.~\ref{fig:phi}) to build a 
light curve template of long-period RRLs (see 
Appendix~\ref{sec:append}).

Second, we selected two sub-groups of T2Cs: One on the extension of 
the RRL locus in the $\phi_{31}$-$\log{P}$ diagram (black box in 
the middle panel of Fig.~\ref{fig:phi}), that we name 
``{\it candidate RRLs}'' (15 objects, see Table~\ref{tbl:rrlcand}) and the other
at lower $\phi_{31}$ and longer periods (green box 
in the middle panel of Fig.~\ref{fig:phi}) that we name 
``{\it short-period T2Cs}'' (99 objects).
Note that the boxes have a purely empirical meaning, 
only to separate the two groups. The 
working hypothesis is that the stars in the 
first sub-group are, indeed, long-period RRLs.

\begin{table}
 \footnotesize
 \caption{List of candidate RRab among Bulge T2Cs.}
 \centering
 \begin{tabular}{l l}
 \hline\hline  
\multicolumn{2}{c}{ID (OGLE IV)\tablefootmark{a}} \\
\hline
0022 & 0517 \\
0041 & 0598 \\
0061 & 0636 \\
0062 & 0638 \\
0228 & 0767 \\
0261 & 0797 \\
0297 & 0909 \\
0326 &      \\
\hline
 \end{tabular}
\tablefoot{
 \tablefoottext{a}{The full name is OGLE-BLG-T2CEP-XXXX, where ``XXXX''
 is the ID appearing in the first column.}}
 \label{tbl:rrlcand}
 \end{table}

\begin{figure}[htbp]
\centering
% \figurenum{1}
\includegraphics[width=8.5cm]{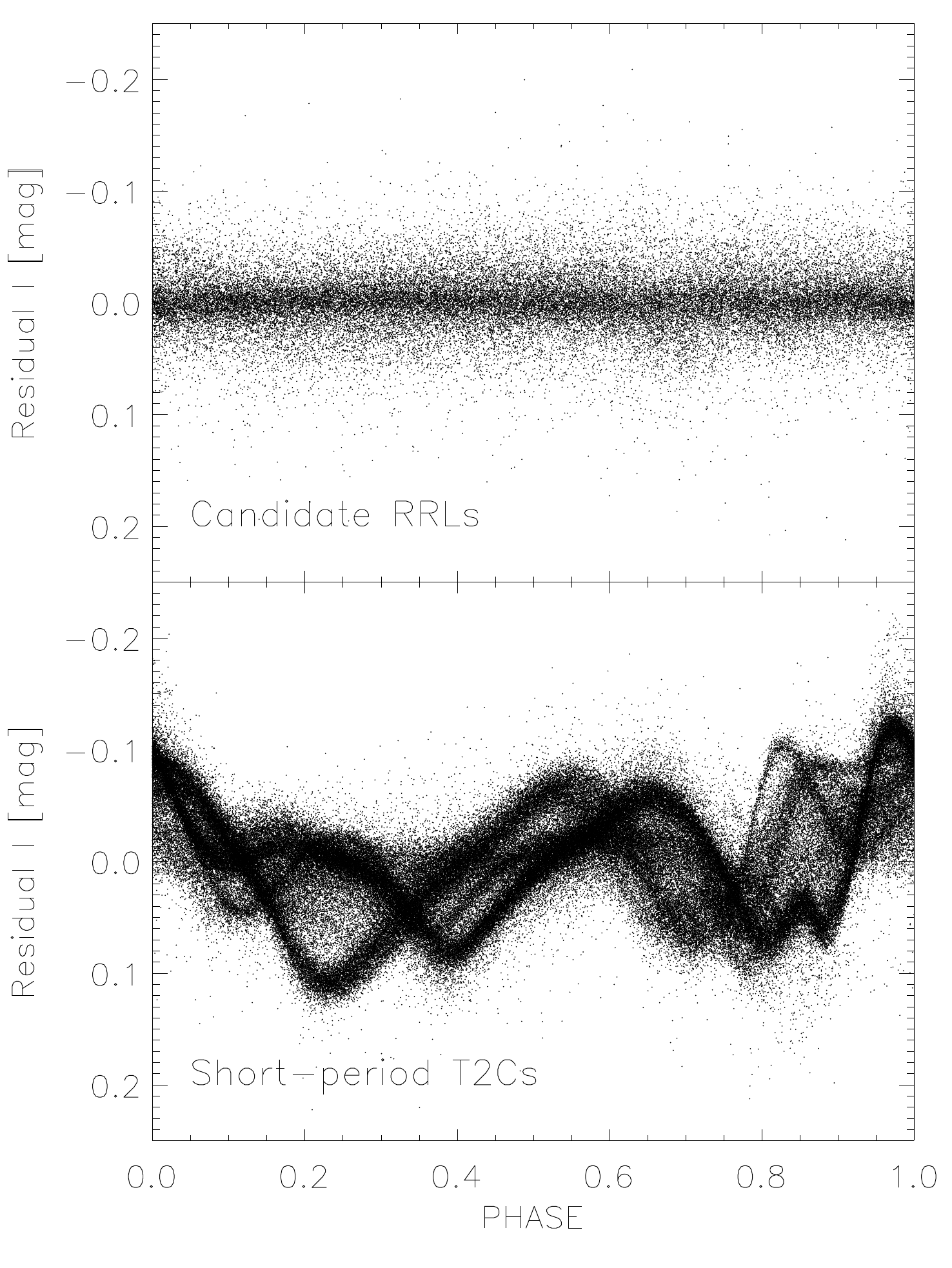}
\caption{Top: Residuals of the $I$-band light curve of all
{\it candidate RRLs} from the template fit. All the 
variables were phased with the period and epoch of maximum
provided by OGLE, meaning that phase 0 is the phase of 
maximum brightness.
Bottom: same as top, but for all the {\it short-period T2Cs}.}
\label{fig:lcvresidual}
\end{figure}

To quantitatively validate this hypothesis,
we fitted the $I$-band light curves of both sub-groups of 
variables with the RRL light-curve templates derived before.
Indeed, we found that the mean standard deviation from the fit
of the light curves of the {\it candidate RRLs}
is 0.016$\pm$0.011 mag. For the second 
sub-group, the mean standard deviation is 0.045$\pm$0.018 mag.
Moreover, a visual inspection of the residuals of 
the light curves from the template fit 
(see Fig.~\ref{fig:lcvresidual}) reveals that the light curves 
of the {\it candidate RRLs} are much closer to those of 
bona-fide RRLs than the light curves of {\it short-period T2Cs}.
In fact, while the residuals of the {\it candidate RRLs} from
the template fit do not follow any trend with the phase, those 
of the {\it short-period T2Cs} show a clear periodic behaviour, 
although this is not the same for all the stars.
Finally, we found that the candidate RRLs
are placed at the lower edge of the T2C distribution in 
the Bailey diagram (see bottom panel of Fig.~\ref{fig:phi}).

%figura generata con temp_200414

Based on these considerations, we conservatively 
classify V92 as a candidate RRab variable. We point 
out that these classifications --despite being crucial to
understand the evolutionary status of the stars-- are
irrelevant concerning the Period-Luminosity
relations and distance estimates, because RRab and T2Cs do
follow the same relations in the NIR 
\citep{matsunaga06,majaess2010}.

% similtemplate rrl per confrontare le curve di luce: temp_200414

%____________________________________________________________________________
\section{Comparison with Type II Cepheids in other stellar systems}\label{chapt_comparison}

T2Cs are not as numerous as RRLs, but they are 
found in all the regions of the Galaxy (Bulge, Halo,
GGCs), with the exception of the thin disk. They are also 
found in the Magellanic Clouds. 
Since their pulsation properties depend on the population 
(metallicity, age) properties of the host system, it is
useful to adopt several diagnostics to compare the 
properties of T2Cs in \wcen~with those in other environments.

\subsection{Bailey diagram: comparison with the Halo}

The Galactic Halo is the most important component of the Galaxy
concerning its merging history. Thanks to the Gaia mission \citep{gaia_alldr}, 
the investigation in Galactic archaeology is at its peak, with streams
and remnants from merged galaxies being found, elucidating the past 
merging history of the Galaxy 
\citep{helmi2018,myeong2019,kruijssen2020,helmi2020}.
\wcen~itself is, most likely, a former GC of an accreted 
galaxy \citep[either Gaia-Enceladus or Sequoia][]{massari2019,bekki2019},
now orbiting the Galactic halo.

The fact that Halo and GGC T2Cs belong to a different 
population was already suggested by \citet{woolley1966}
and recently confirmed by \citet{wallerstein2018},
based on both kinematics and metal abundance. 
It is therefore interesting to compare the properties of \wcen~T2Cs
with those in the Halo.
As a diagnostic, we adopt the Bailey diagram, independent of 
distance and reddening.
% colour-colour diagrams cannot be derived for Halo 
% T2Cs since we would need well-sampled light curves in three passbands.

\begin{figure}[htbp]
\centering
% \figurenum{1}
\includegraphics[width=9cm]{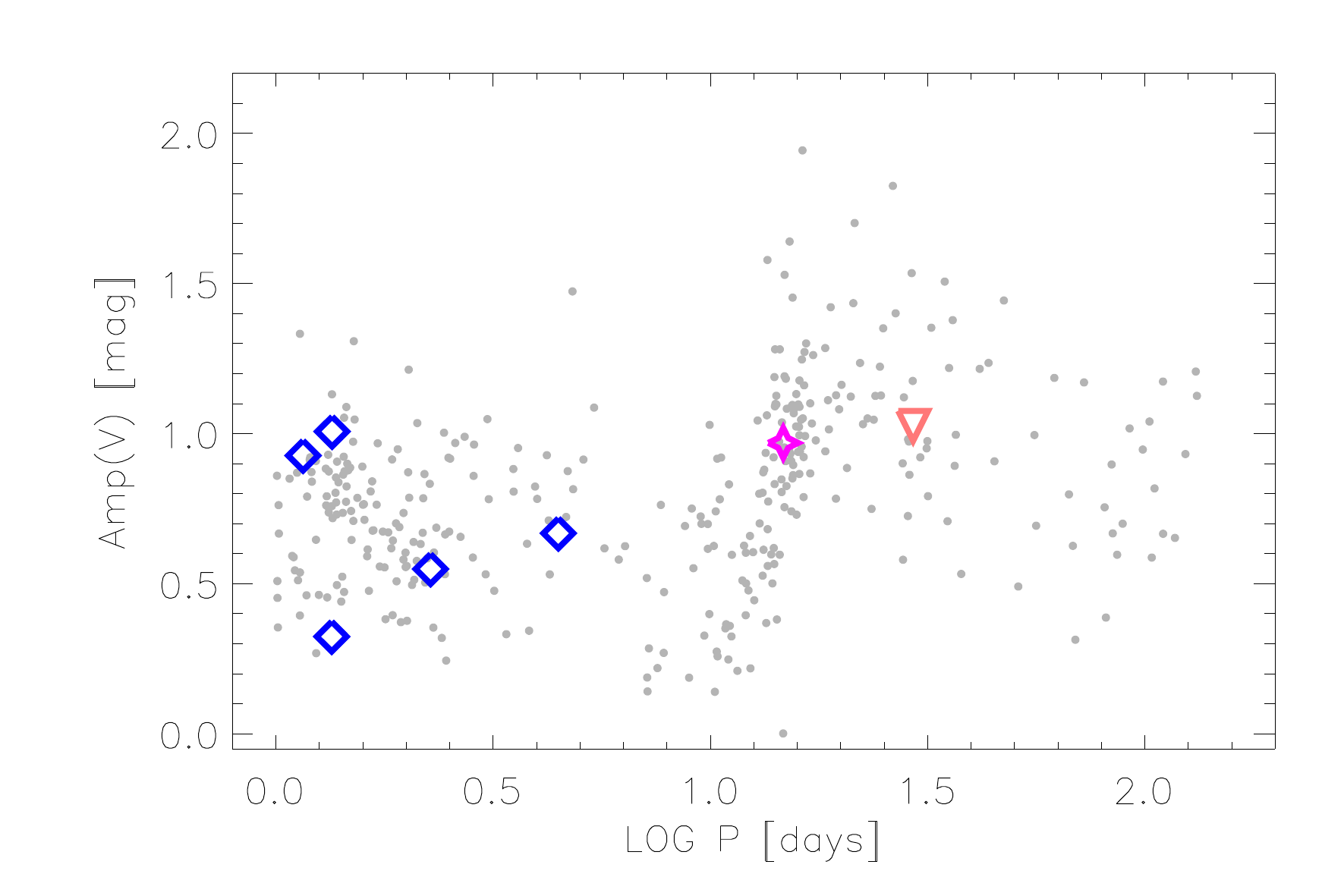} %IDLWorkspace/Default/Temp/temp_181213.pro
\caption{Bailey diagram of T2Cs in the Halo (grey points), 
and in \wcen~(same symbols as in Fig.\ref{fig:basti}, plus
magenta star for the WV and orange upside-down triangle for the RVT.)}
\label{fig:bailey_field}
\end{figure}

Figure~\ref{fig:bailey_field} displays the optical ($V$-band) Bailey diagram
of 362 Halo T2Cs and \wcen~T2Cs. We adopted the catalogue of Cepheids
within Gaia DR2 published by \citet{ripepi2019}. However, we 
both complemented and corrected this list, by comparing their
classification with other surveys and literature data, namely,
ASAS \citep{pojmanski1997}, ASASSN \citep{shappee2014,jayasinghe2019},
GCVS \citep{samus2017} and \citet{warren1976}. 
We provide in Table~\ref{tab:newcepheids} the list of T2Cs that
we added to the \citet{ripepi2019} sample and those for which
we have changed the classification.

\begin{table}   
\scriptsize   
\caption{Complements and changes to the \citet{ripepi2019} T2C catalogue.}   
\label{tab:newcepheids}   
\centering   
\begin{tabular}{l l l l l}   
\hline   
\hline   
Name & Gaia DR2 ID & class (R19) & new class & Ref.\tablefootmark{a} \\   
\hline   
XX Vir & 3640760901131104256 & N/A & BLH & 2         \\
VZ Aql & 4205497393488435200 & N/A & BLH & 2         \\
V439 Oph & 4472449191647245184 & N/A & BLH & 2       \\
BL Her & 4527596850906132352 & N/A & BLH & 3           \\
V446 Sco & 4037438101994829312 & N/A & WV & 2        \\
AL CrA & 4037674673147744384 & N/A & WV & 2          \\
V564 Sgr & 4042147241577658880 & N/A & WV & 1          \\
V1834 Sgr & 4045437774995118976 & N/A & WV & 2       \\
V1303 Sgr & 4052014091545300480 & N/A & WV & 2       \\
V1185 Sgr & 4052361842043219328 & N/A & WV & 1         \\
V802 Sgr & 4073100869046439040 & N/A & WV & 5      \\
V554 Oph & 4117307863590649600 & N/A & WV & 1          \\
BH Oph & 4484791347109688832 & N/A & WV & 2          \\
HQ Car & 5254665166975458944 & N/A & WV & 2          \\
MR Ara & 5954403987593491584 & N/A & WV & 1            \\
AL Vir & 6303152720661307648 & N/A & WV & 3            \\
BO Tel & 6643297500393311616 & N/A & WV & 2          \\
V347 CrA & 6726052960315592576 & N/A & WV & 1          \\
V383 Sgr & 6736147782729787264 & N/A & WV & 2        \\
SZ Mon & 3112344688094507136 & N/A & RVT & 3           \\
MZ Cyg & 1964010169902699648 & WV & RVT & 2     \\
IU Cyg & 2035402872974695936 & --- & RVT & 2      \\
V1831 Sgr & 4048985899093631616 & N/A & RVT & 2      \\
TZ Ser & 4161479334488856320 & WV & RVT & 1       \\
EP Mus & 5855676944429471872 & WV & RVT & 2     \\
CQ Sco & 5957918469109353216 & N/A & RVT & 2         \\
RX Lib & 6241789522177233664 & N/A & RVT & 2         \\
ET Oph & 4111880369315900032 & RVT & RVT/DCEP & 4/2     \\
\hline   
\end{tabular}   
\tablefoot{\tablefoottext{a}{Reference for the new classification
1: ASAS, 
2: ASASSN, 
3: GCVS, 
4: Gaia DR2 \citep{ripepi2019}, 
5: \citet{warren1976}}}   
\end{table}   

We note that the Halo field hosts T2Cs with periods longer
than 100 days. This is a remarkable difference compared to T2Cs in all GGCs 
\citep[V16 in NGC~6569 with a period of 87.5 days,][]{Clement01} and---as we will show in 
Section~\ref{chapt_comparison_bulge}---with those in the Galactic Bulge  
(the longest period is 84.8 days for T2Cs in the outer Bulge 
\citep{soszynski2017} and 93.5 days for T2Cs in the inner Bulge 
\citep{braga2019b}). This is further evidence that T2Cs in the Halo
belong to a different population than T2Cs in GGCs. We also note that the 
four short-period BLHs of \wcen~are placed at the 
edges of the distribution of field T2Cs. 
More precisely, V43 and V60 are at the high-amplitude edge,
while V61 and V92 are placed around the low-amplitude edge. 

%____________________________________________________________________________
\subsection{Amplitude ratios: comparison with the Bulge}\label{chapt_comparison_bulge}

Within the OGLE survey, more than 1,000 T2Cs were detected in the 
Galactic Bulge \citep[][and further addenda]{soszynski2018}.
Although both $V$- and $I$-band time series are available, only
$Amp(I)$ were published. Therefore, we have downloaded the $V$-band 
time series and derived $Amp(V)$, that we publish in 
Table~\ref{tab:vampl}. Note that, for RVTs showing 
alternating deep and shallow minima, we folded the 
light curves at their pulsation
period (that is, the period between two relative minima). This means
that their folded light curves display a wide dispersion
around the minimum and our $Amp(V)$ estimates for these stars 
are an average between their minimum amplitude 
(shallow minimum-to-maximum magnitude difference) 
and maximum amplitude (deep minimum-to-maximum magnitude difference).
To compare the Bailey diagrams also in the NIR, we have adopted the 
$Amp(K_s)$ of the same variables obtained from 
VVV data \citep{braga2018b}. 

\begin{table}
 \footnotesize
 \caption{$Amp(V)$ of Bulge T2Cs.}
 \centering
 \begin{tabular}{l c}
 \hline\hline  
 ID (OGLE IV)\tablefootmark{a} & Amp($V$) \\
  & mag \\
\hline
 0001 & \ldots  \\
 0002 & \ldots  \\
 0003 & \ldots  \\
 0004 & \ldots  \\
 0005 & \ldots  \\
 0006 & 0.894$\pm$0.090  \\
 0007 & 0.838$\pm$0.069  \\
 0008 & \ldots  \\
 0009 & 0.561$\pm$0.036  \\
 0010 & \ldots  \\
\hline
 \end{tabular}
\tablefoot{Only the first 10 of the 1068 lines of 
the table are shown. The full table is
 shown in the machine-readable version of the paper.\\
 \tablefoottext{a}{The full name is OGLE-BLG-T2CEP-XXXX, where ``XXXX''
 is the ID appearing in the first column.}}
 \label{tab:vampl}
 \end{table}

\begin{figure}[htbp]
\centering
% \figurenum{1}
\includegraphics[width=8cm]{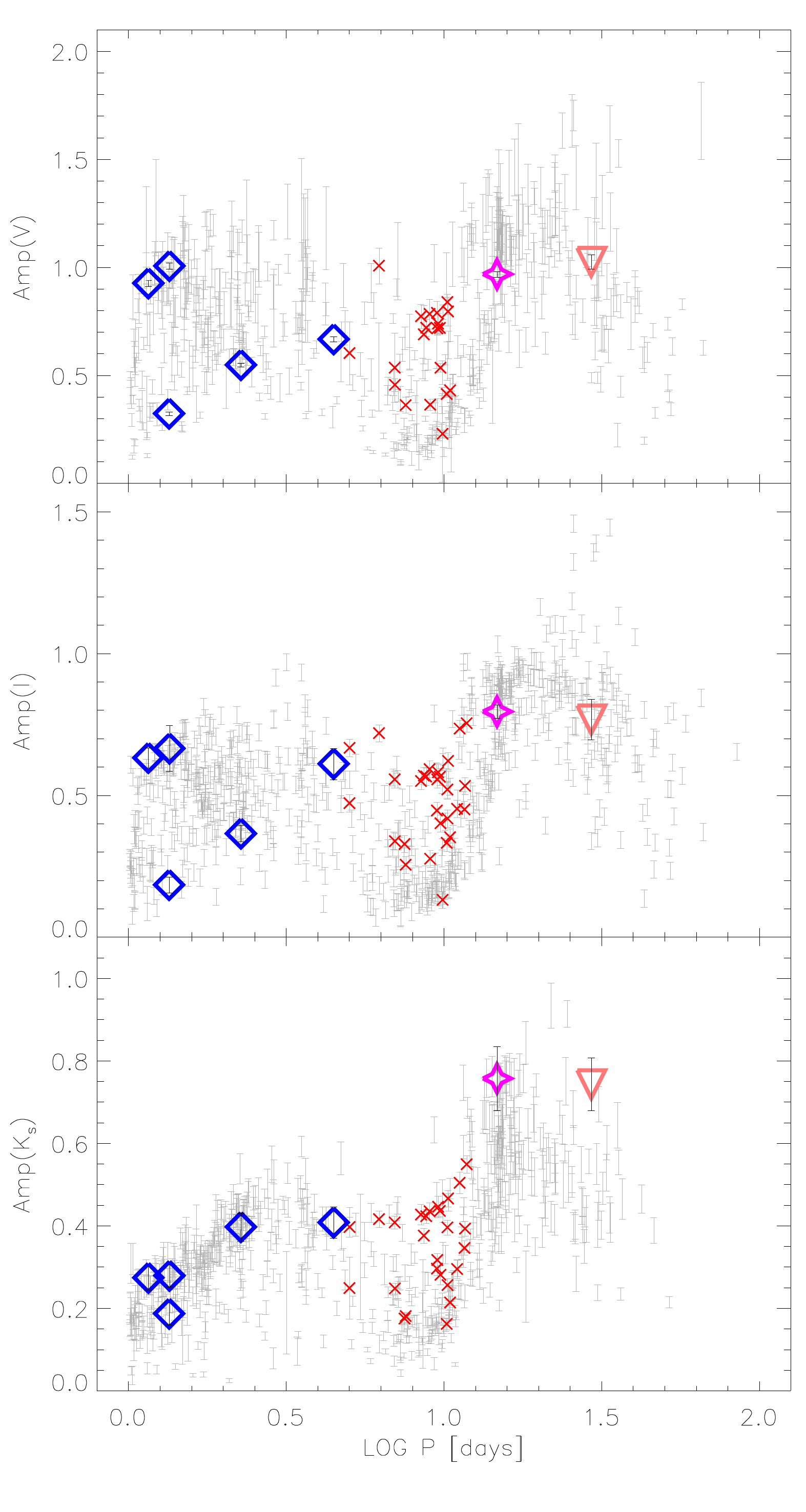}
\caption{Top: Bailey diagram of Bulge and \wcen~T2Cs in
the $V$ band. Grey: Bulge BLHs; red crosses: Bulge pWVs; 
other symbols are for \wcen~T2Cs and have 
the same meaning as in Figs.~\ref{fig:basti} 
and~\ref{fig:bailey_field}. 
The uncertainties of the light curve amplitudes 
of \wcen~T2Cs are displayed as black error bars.
Middle: same as top, but for the 
$I$ band. Note that the 
OGLE catalogues do not provide the uncertainties on the $Amp(I)$,
therefore we assumed an uncertainty of 0.05 mag.
This is a conservative assumption, since 
the median of the uncertainties on $Amp(V)$ is 0.052 mag, 
and the $I$-band time series of OGLE have about one 
order of magnitude more points than the $V$-band time series.
Bottom: same as top, but for the $K_s$ band.}
\label{fig:bailey_bulge}
\end{figure}

The Bailey diagrams in all three
bands are displayed in Fig.~\ref{fig:bailey_bulge}. 
The two optical ones are quite similar.
In fact, BLHs display a very shallow and high-dispersion increase
in amplitude from 1 to $\sim$3.2 days. In the $K_s$ band, the increase 
is not only steeper, but also displays a smaller dispersion.
Starting from $\sim$3.2 days up to the whole period range of T2Cs, 
all three Bailey diagrams show a minimum at $\sim$8 days and 
a subsequent increase in amplitude, until reaching another
maximum at $\sim$20 days, which is also the threshold between
WVs and RVTs. This feature was recently discussed, from a theoretical 
persective, in Bono et al. (2020, submitted).
At longer periods, the behaviour is not clear
due to a large dispersion, but a general decrease in amplitude 
is observed in all three bands.

We point out that the short-period BLHs of \wcen~are at the lower
and upper edge of the distribution in the Bailey diagrams,
especially the optical ones. This is the same 
behaviour that we observed when comparing to 
Halo T2Cs. In passing, we note that the peculiar WV stars 
(pWVs) are, at fixed period, brighter than canonical WV and 
are thought to belong to binary systems 
\citep{soszynski08c,pilecki17,pilecki2018}.
The pWVs in the Bailey diagram appear to have optical amplitudes 
($Amp(V)$, $Amp(I)$ ) that are either similar or larger than 
those of canonical WVs. In the NIR the trend is not so clear.

% figure generate con temp_180425
% cp /media/vittorio/Volume/Surveys/VVV/data/bailey_amplratio_ogle4_3* .
%%%%%%%%%%%% Fig 3 %%%%%%%%%%%%%%%%

\begin{figure}[htbp]
\centering
% \figurenum{1}
\includegraphics[width=8cm]{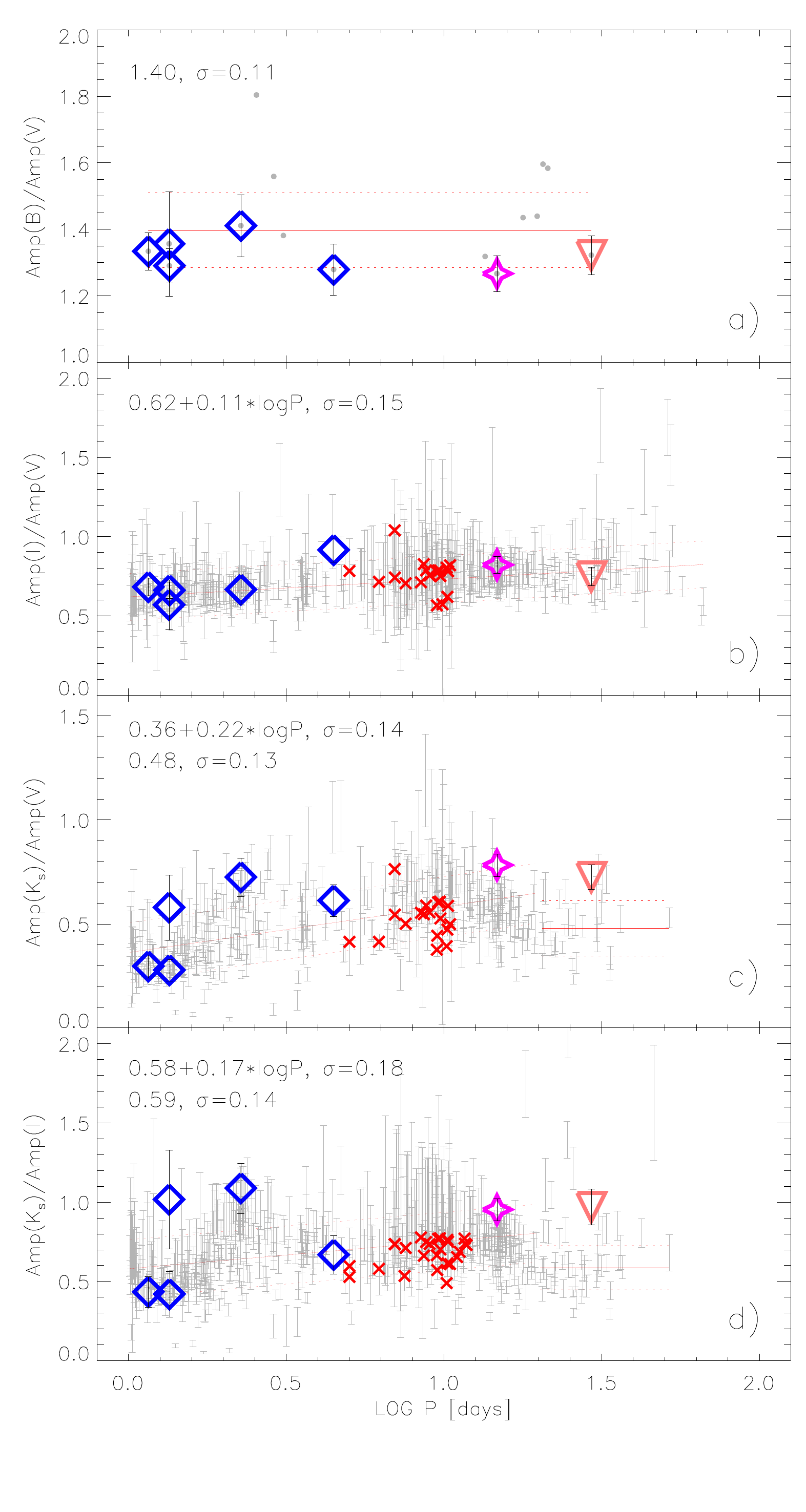}
\caption{Panel a: $B$ over $V$ amplitude ratios of GGC T2Cs.  Black: T2Cs from NGC~6388 
and NGC~6441; other symbols have the same meaning as in Figs.~\ref{fig:basti} 
and~\ref{fig:bailey_field}. 
The solid red line denotes the average while the dotted lines denote the 
1$\sigma$ dispersion.
Panel b: $I$ over $V$ amplitude ratios of Bulge and \wcen~T2Cs. Here, the red
line displays a linear fit. The symbols have the same 
meaning as in Fig.~\ref{fig:bailey_bulge}. Panel c: same as panel b, 
but for $K_s$ over $V$ amplitude ratios. Here, the red line displays a linear
fit for logP$<$1.3 and the average for logP$>$1.3. Panel d: same as panel b, 
but for $K_s$ over $I$ amplitude ratios. Here, the red line displays a linear
fit for logP$<$1.3 and the average for logP$>$1.3.}
\label{fig:amplratio}
\end{figure}

We adopted $Amp(V)$, $Amp(I)$ and $Amp(K_s)$ to derive the 
amplitude ratios of T2Cs (see Fig.~\ref{fig:amplratio}). 
$Amp(I)/Amp(V)$ shows a clear linear trend with the period, and  
it is also interesting to notice that the zero point (0.62) is 
almost identical to the $Amp(I)/Amp(V)$ amplitude ratio of RRLs 
\citep[0.63$\pm$0.01]{braga16}. 
It would be tempting to adopt $Amp(I)/Amp(V)$ to separate
RRLs and T2Cs, because of their different behaviour (RRLs 
display a constant $Amp(I)/Amp(V)$). However, 
the intrinsic dispersion of the trend and the uncertainties on
the amplitudes are larger than the difference of $Amp(I)/Amp(V)$
between RRLs and T2Cs, especially at short periods ($<$3 days),
where RRLs disguised as T2Cs are to be searched.
The ratios involving the $K_s$-band 
amplitudes display a less clear behaviour, with
BLHs+WVS and RVTs following different trends. 
More precisely, both $Amp(K_s)/Amp(V)$ and 
$Amp(K_s)/Amp(I)$ ratios increase with period in the BLHs 
range, approach a steady value and then decrease again 
in the 16-20 days range. However, we conservatively fit 
their distribution with a simple linear fit, also
because the dispersion is quite large compared to 
the average ratio.

Unfortunately, in the VVV data many RVTs are saturated
\citep{bhardwaj17c,braga2018b} therefore we have less points than for 
$Amp(I)/Amp(V)$. The ratios of RVTs in the 
$K_s$-band do not show any clear dependence 
on period and they are lower than those of WVs. 
We ascribe the different behaviour of the NIR 
amplitude ratios of RVTs to the presence 
of circumstellar dust. The long-wavelength
excess light could explain the fact that, in the NIR,
amplitudes are smaller than expected.

Our referee noted that the 
dispersion in the optical/NIR amplitude
ratios is larger than intrinsic errors and might 
be caused by a possible dependence on metal content.

\wcen~T2Cs follow, within 2$\sigma$, the trends that we 
have found, especially for the $Amp(I)/Amp(V)$ ratio, 
that has the tightest relation. 

% We also note that Bulge pVWs are mostly placed 
% below the relation for the $K_s$-band amplitude ratios,
% despite having amplitudes larger, on average, than
% WVs of the same period 
% (see the last panel in Fig.~\ref{fig:bailey_bulge}).
% This might be due to hot and compact (e.g., white
% dwarfs) companions, if pWVs are indeed binary stars.

% figure generate con temp_180425
% cp /media/vittorio/Volume/Surveys/VVV/data/bailey_amplratio_ogle4_3* .
%%%%%%%%%%%% Fig 3 %%%%%%%%%%%%%%%%

\section{Period-Luminosity relation}\label{chapt_distance_omega}

T2Cs are distance indicators that are not so
widely used as CCs and RRLs,
since they are not as numerous. For this reason, the calibration
of their PL still lags that of RRLs and CCs. 
However, T2Cs in GGCs already proved to be crucial for understanding the 
difference between Population I and Population II stars \citep{baade56}.

\subsection{The NIR PLs of T2Cs in GGCs}\label{chapt_comparison_cluster}

Although not all clusters host T2Cs, these variables are 
quite iconic for GGCs and were named "cluster cepheids"
until the 1950s. 

\citet{matsunaga06} performed an analysis 
of the NIR PL relations of the T2Cs in 23 GGCs. 
Our aim is to complement their sample and 
update the relations.
For \wcen, they had data only for V1, V29 
and V48. Moreover, they adopted E(\bmv) from the 
Harris catalogue of GGCs \citep{harris96} and derived the 
true distance moduli (DM$_0$) from the
$M_V$-[Fe/H] relation of HB stars, adopting
$V_{HB}$ and [Fe/H] from the Harris catalogue. However, not only 
this catalogue contains heterogeneous data, mostly 
based on optical investigations, but the 
$M_V$-[Fe/H] relation is prone to both non-linearity 
and evolutionary effects \citep{caputo00}.
Moreover, several GGCs are located in 
the Galactic Bulge and recent NIR investigations 
provide more reliable estimates. 

Therefore, we have collected DM$_0$ 
and E(\bmv) of GGCs from more recent literature
to derive the absolute magnitudes. 
When possible, we have favoured papers providing distance estimates based on 
PL relations of variable stars and on NIR data. 
This means that, in the end, our absolute magnitudes are slightly
different compared to those of \citet{matsunaga06}. 
The degree of homogeneity is not complete, 
but is higher than that of the Harris catalogue. 
All references are listed in Table 7~\ref{tbl:ggc_dmod_ebv}.

\begin{table}
\footnotesize
\caption{Distance moduli and reddening of GCs.}
\label{tbl:ggc_dmod_ebv}
\centering
\begin{tabular}{l l l l}
\hline
\hline
GC & $\mu$ & E(\bmv) & Ref.\tablefootmark{a} \\
  & mag & mag &  \\
\hline
 NGC 2808  &  15.04 & 0.17  &  1    \\     
 NGC 5272  &  15.07 & 0.01  &  2,3  \\
 NGC 5904  &  14.35 & 0.035 &  2,4  \\  
 NGC 5986  &  15.10 & 0.28  &  3    \\  
 NGC 6093  &  15.01 & 0.18  &  3    \\
 NGC 6218  &  13.43 & 0.19  &  3    \\     
 NGC 6254  &  13.22 & 0.28  &  3    \\     
 NGC 6256  &  14.79 & 1.20  &  5    \\
 NGC 6266  &  14.11 & 0.47  &  5    \\
 NGC 6273  &  14.58 & 0.40  &  5    \\     
 NGC 6284  &  15.93 & 0.28  &  3    \\
 NGC 6325  &  14.51 & 0.91  &  3    \\     
 HP 1      &  14.17 & 1.18  &  6    \\     
 Terzan 1  &  14.13 & 1.99  &  6    \\     
 NGC 6402  &  14.85 & 0.60  &  3    \\    
 NGC 6441  &  15.65 & 0.52  &  5    \\    
 NGC 6453  &  15.15 & 0.69  &  6    \\    
 NGC 6569  &  15.40 & 0.49  &  5    \\
 NGC 6626  &  13.73 & 0.40  &  3    \\
 NGC 6749  &  14.45 & 1.50  &  7    \\
 NGC 6779  &  15.03 & 0.18  &  8    \\    
 NGC 7078  &  15.13 & 0.09  &  2,4  \\    
 NGC 7089  &  15.09 & 0.06  &  9,3 \\
\hline
\end{tabular}
\tablefoot{\tablefoottext{a}{When two references are given, the first is for 
$\mu$ and the second for E(\bmv). \\ 
1: \citet{kunder2013a}, 
2: \citet{sollima06}, 
3: \citet{harris96}, 
4: \citet{ferraro1999}, 
5: \citet{valenti2007}, 
6: \citet{valenti2010},
7: \citet{kaisler1997},
8: \citet{ivanov2000},  
9: \citet{lazaro2006}.}}
\end{table}

Finally, we have added V43, V60, V61 and 
V92 to the sample of \citet{matsunaga06}, and replaced 
their mean magnitudes of
V1, V29 and V48 with our own values, since our 
light curves are better sampled.

% Note that we have used, for RVTs, the half-period, that
% is defined as the time elapsed between two relative minima.
% In the case of GGC RVTs, relative and absolute minima are the same,
% because they do not show alternating deep and shallow minima
% like field T2Cs.

%%%%%%%%%%%% Fig 3 %%%%%%%%%%%%%%%%

% figura generata con cmd_bvi2
% cp ../t2c_ggcs_pl_jhk2_t2c-eps-converted-to.pdf .
\begin{figure}[htbp]
\centering
% \figurenum{1}
\includegraphics[width=7cm]{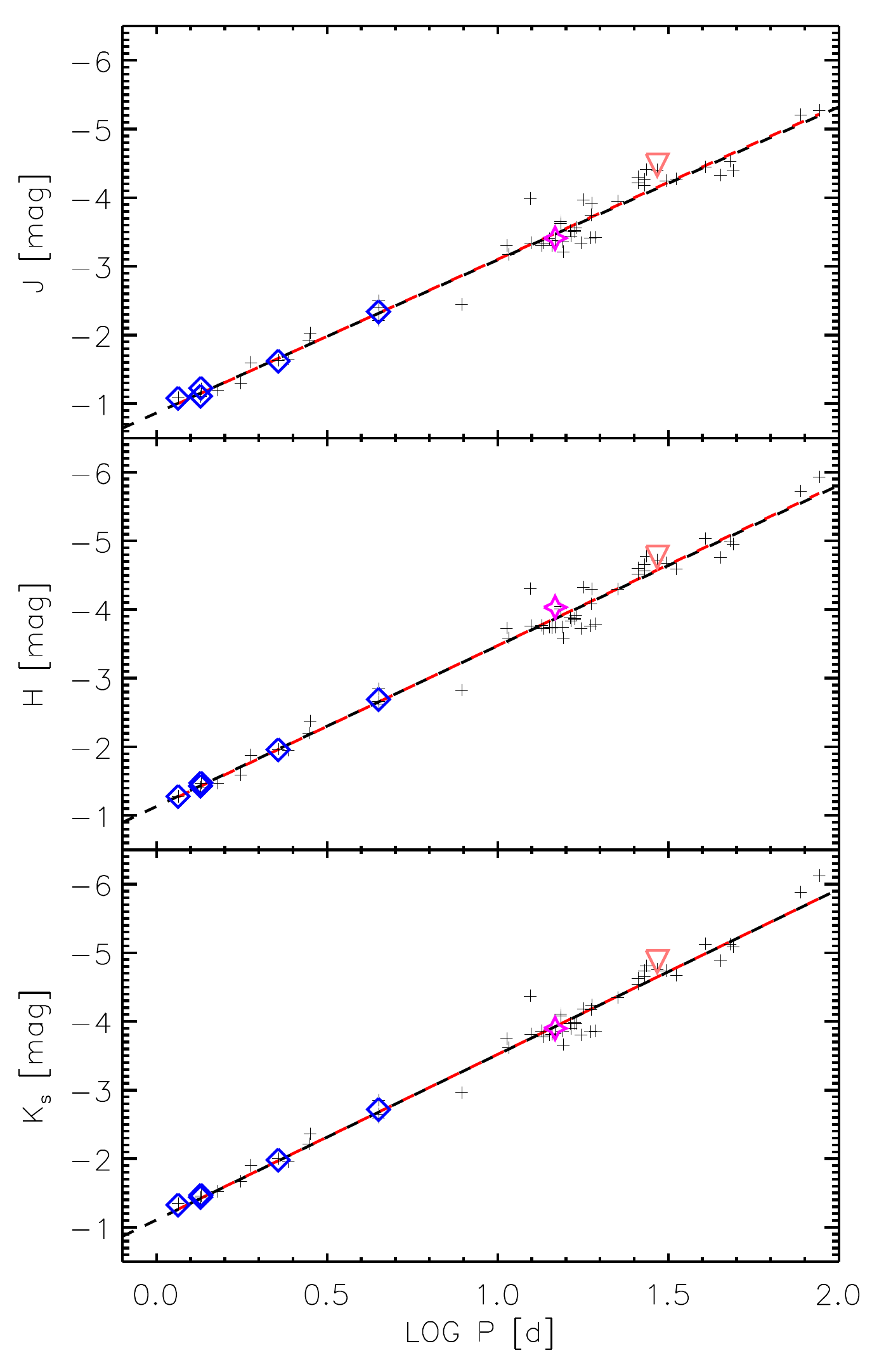}
% \vspace*{1.0truecm}
\caption{Top: $J$-band PL relation of T2Cs in GGCs. 
Symbols for \wcen~are the same as in Figs.~\ref{fig:basti} 
and~\ref{fig:bailey_field};
T2Cs from other GGCs are displayed as black crosses.
The dashed black line displays the PL found 
by \citet{matsunaga06}, while the dashed red line
displays our own PL.
Middle: same as top but for the $H$-band PL relation.
Bottom: same as top but for the $K_s$-band PL relation.}
\label{fig:t2cep_ggcs}
\end{figure}

The updated PL relations that we have derived 
agree very well with those found by \citet{matsunaga06}
(see Table~\ref{tbl:ggc_matsunaga}).
By comparing our Figure~\ref{fig:t2cep_ggcs} with 
Figure~3 in \citet{matsunaga06}, 
we note that the T2Cs in \wcen~that we have in common
(V1, V29 and V48), follow even more
closely the NIR PL relations. 
The only relevant difference is the position of 
\ngc{6256} V1: in our figure, it is $\sim$0.6 mag
brighter in all bands, meaning an offset larger than
3$\sigma$.

\begin{table}
\footnotesize
\caption{Empirical NIR PL relations of T2Cs in GGCs.}
\label{tbl:ggc_matsunaga}
\centering
\begin{tabular}{l c c c}
\hline
\hline
Band & a & b & $\sigma$ \\
  & mag & mag & mag \\
\hline
\multicolumn{4}{c}{---\citet{matsunaga06}---} \\
$J$ &   --0.86$\pm$0.08   &   --2.23$\pm$0.07   &   \ldots \\
$H$ &   --1.13$\pm$0.07   &   --2.34$\pm$0.06   &   \ldots \\ 
$K_s$ &   --1.11$\pm$0.07   &   --2.41$\pm$0.06   &   \ldots \\
\multicolumn{4}{c}{---Our coefficients---} \\
$J$ &   --0.86$\pm$0.06   &   --2.23$\pm$0.05   &   0.17 \\
$H$ &   --1.11$\pm$0.07   &   --2.36$\pm$0.06   &   0.19 \\ 
$K_s$ &   --1.12$\pm$0.05   &   --2.40$\pm$0.05   &   0.16 \\
\hline
\end{tabular}
\tablefoot{$M_X$ = a + b$\cdot\log{P}$.}
\end{table}

We point out that NGC 6256 is
a Bulge GGC, affected by high extinction, with
a large uncertainty: E(\bmv) ranges from 
0.84 \citep{harris96} to 1.66 mag \citep{Schlegel98}.
Moreover, there is solid empirical evidence 
to believe that the reddening in NGC 6256 is, most probably, differential
\citep{valenti2007}. We have also checked that 
V1 is not blended in the IRSF $JHK_s$ images of \citet{matsunaga06}.
In the end, we cannot exclude the 
possibility that \ngc{6256} V1 is a 
pWV. This is a kind of object that is slightly 
brighter than WVs and that is likely 
to belong to a binary system \citep{soszynski08c}.
This would be the first pWV found in a GGC, since pWVs have already 
been identified in the Magellanic Clouds 
\citep{soszynski08c,soszynski10c}, in the field
\citep[$\kappa$ Pav,][]{matsunaga2009a}
and in the Bulge \citep{soszynski2017}. 

% Since pWVs are likely 
% binary systems, we put this in correlation with the fact that
% \ngc{6256} is a post core-collapse GGC, and one of the few 
% hosting T2Cs XXX c'entra veramente qualcosa???? e' 
% piu probabile trovare binarie in questo caso??? XXX.

% %_______________________________________________________________________________
\subsection{The PL transition from RRLs and T2Cs and distance determination}\label{chapt_t2c_pl}

\begin{figure}[htbp]
\centering
\label{fig:fig_pl_opt}
\includegraphics[width=8.5cm]{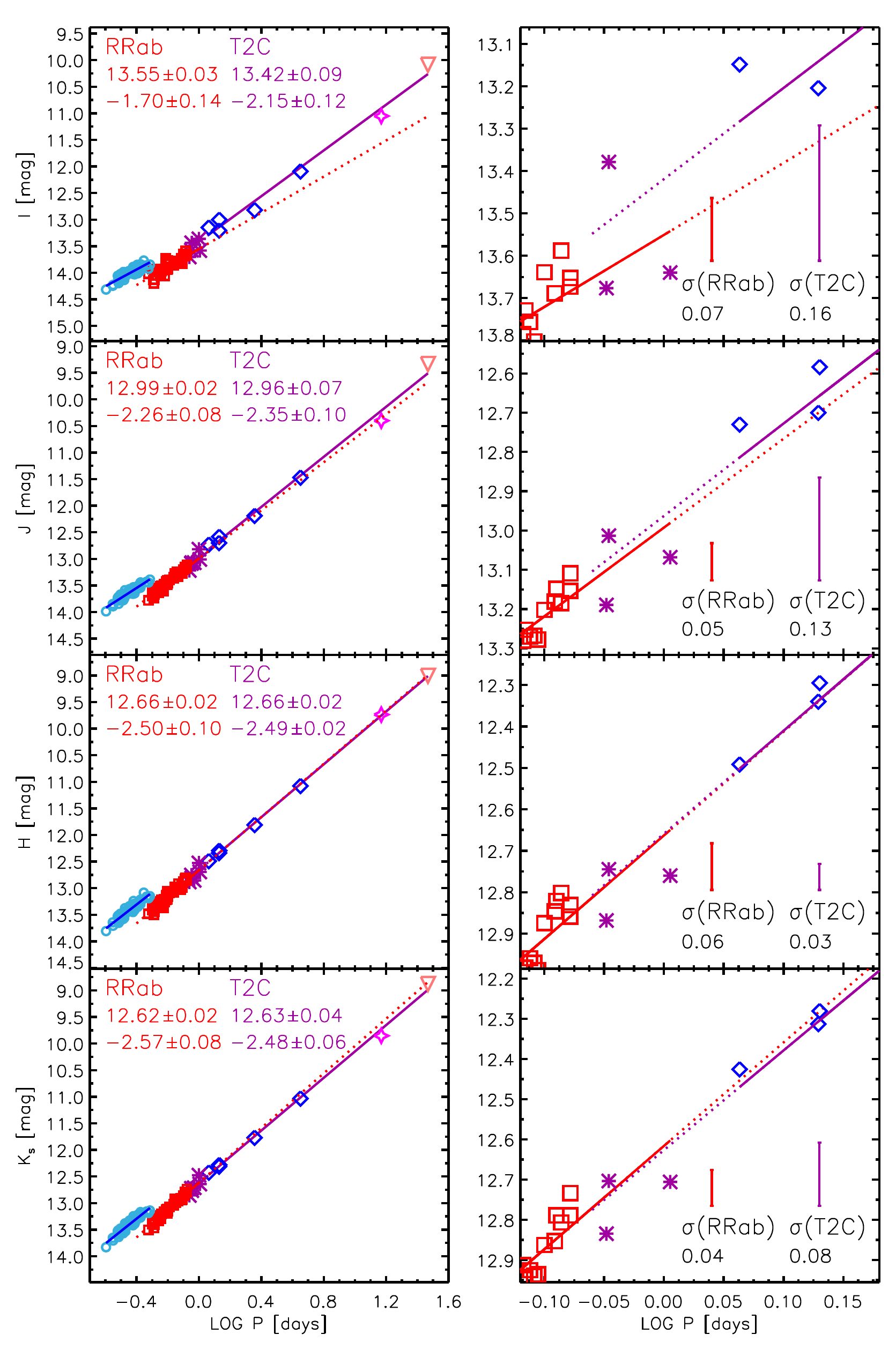}
\caption{$IJHK_s$ PL relations of RRLs and T2Cs in \wcen. The symbols
have the same meaning as in Figs.~\ref{fig:basti} 
and~\ref{fig:bailey_field}. Left panels display
both RRLs and T2Cs, while right panels are a close-up on the 
period range of the transition between long-period RRab and BLHs.
Red and purple lines show the PL relation of RRab and
T2Cs, respectively. The dotted lines show the extension at
longer/shorter periods of the RRab/T2Cs PL relations, respectively.
The dispersions of the relations are displayed as bars in the 
lower-right corner of the right panels.}
\end{figure}

Figure~\ref{fig:fig_pl_opt} displays the PL relations of RRLs and 
T2Cs of \wcen, from the $I$ to the $K_s$ band. Given that RRLs do not follow tight 
PL relations in passbands bluer than $R$ \citep{catelan04,braga15}, 
we do not display the $B$ and $V$ PL relations. The $R$-band PL 
relations are not discussed, because the light curves 
are hampered by poor/modest sampling. Theory and observations indicate that the PL
relations of RRLs are mildly affected by metallicity \citep{marconi15}, while T2Cs
are not \citep{dicriscienzo07,lemasle15}, we decided to rescale the magnitudes of all
\wcen~RRLs to the value they would have at the average metallicity of the sample.
To do so, we have 
adopted the theoretical metallicity coefficients of the PLs
by \citet[][see their Table 6]{marconi15} and the new spectroscopic
[Fe/H] estimates by \citet{magurno2019}. The rescaled 
magnitudes in a generic filter $X$ ($X_{Fe}$) have been determined as 

$$ X_{Fe} = X - c \cdot ([Fe/H] - \langle[Fe/H]\rangle) $$

where c is the metallicity coefficient and $\langle[Fe/H]\rangle$ is
the mean metallicity---calculated as $\log{(\langle 10^{[Fe/H]}\rangle)}$---of \wcen~RRLs.
These rescaled magnitudes were adopted to derive 
the PL relations of RRLs displayed in Fig.~\ref{fig:fig_pl_opt}

% figure generate con pl_plw_omegacen.pro
% cp ~/Documenti/Science/RRLyr/Omega_Cen/PLW/PL_empirical_rrlt2c_xxx3_IJHK_clip28-eps-converted-to.pdf .

We note that, with the only exception of the $H$-band,
V29 (the WV) is always underluminous compared to the T2C 
relation, while V1 (the RVT) is always overluminous. If the PL
relations of T2Cs were affected by metallicity in the same way 
as RRLs, we would expect an opposite behaviour,
based on the metal abundances by
\citet{gonzalez1994b}. This is further evidence that 
the effect of metallicity on the PL relations of T2Cs is 
negligible.

We notice that in the $JHK_s$ bands, the 
long-period RRL stars follow both the RRab and the T2C relations.
Most importantly, the coefficients of the RRab and T2C relations
are the same within 1$\sigma$ in the $J$ and $K_s$ bands, and 
are practically identical in the $H$ band. This means that 
RRab and T2Cs obey a common NIR PL relation, as already 
suggested by \citet{majaess2010}.

\begin{table}
\footnotesize
\caption{Empirical Optical and NIR PL relations of T2Cs in \wcen.}
\label{tbl:empirical_pl}
\centering
\begin{tabular}{l c c c}
\hline
\hline
Band & a & b & $\sigma$ \\
 & mag & mag & mag \\
\hline
%       $B$   & 14.563$\pm$ 0.204 & --1.821$\pm$ 0.267 &  0.364 \\
%       $V$   & 14.111$\pm$ 0.153 & --2.023$\pm$ 0.199 &  0.272 \\
      $I$   & 13.419$\pm$ 0.090 & --2.151$\pm$ 0.117 &  0.160 \\
      $J$   & 12.963$\pm$ 0.074 & --2.354$\pm$ 0.096 &  0.131 \\
      $H$   & 12.659$\pm$ 0.018 & --2.485$\pm$ 0.023 &  0.031 \\
      $K_s$ & 12.627$\pm$ 0.044 & --2.479$\pm$ 0.058 &  0.079 \\
\hline
\end{tabular}
\tablefoot{$X$ = a + b$\cdot\log{P}$.}
\end{table}

To estimate the DM$_0$ of \wcen, we adopt both our own
empirical calibration of the NIR PL relation based on GGC T2Cs
(Table~\ref{tbl:ggc_matsunaga}) and, as 
a comparison, that by \citet{matsunaga06}.\footnote{ 
The coefficients in Table~\ref{tbl:ggc_matsunaga} 
were obtained by including \wcen~ T2Cs. 
In principle, for a rigorous estimate of \wcen~distance 
we should re-derive the coefficients of 
the NIR PL relations by removing \wcen~T2Cs from the sample discussed
in Section~\ref{chapt_comparison_cluster}. However, 
we have checked that, within the errors, 
the coefficients are the same, either including or
neglecting \wcen~T2Cs from the sample.}
Finally, we derived the 
distance moduli, listed in Table~\ref{tbl:distance}.
  
% PLZ_empirical5_DM_clip28.tex
\begin{table}
\footnotesize
\caption{True Distance Moduli to \wcen}
\label{tbl:distance}
\centering
\begin{tabular}{l c c c}
\hline
\hline
Band & DM$_0$ & err & $\sigma$ \\
     & mag& mag & mag \\
\hline
% %------- mia calibrazione (omega cen escluse)------
% $J$           & 13.666 & 0.081 &  0.138 \\
% $H$           & 13.635 & 0.077 &  0.070 \\
% $K_s$         & 13.649 & 0.072 &  0.077 \\
% $mean_{JHKs}$ & 13.650 & 0.077 &  0.016 \\
% %------- calibrazione Matsunaga ------
% $J$           & 13.656 &  0.081 &  0.138 \\
% $H$           & 13.649 &  0.077 &  0.085 \\
% $K_s$         & 13.658 &  0.072 &  0.081 \\
% %------- mia calibrazione (omega cen incluse)------
$J$           & 13.656 &  0.066 &  0.137 \\
$H$           & 13.640 &  0.073 &  0.075 \\
$K_s$         & 13.663 &  0.061 &  0.084 \\
$mean_{JHKs}$ & 13.649 & 0.067 &  0.008 \\
\hline
\end{tabular}
\end{table}

By following the referee's suggestion, we have obtained the 
$K_s$,$J-K_s$ Period-Wesenheit (PW) relation, adopting the total-to-selective
extinction ratio both by \citet[][$\frac{AK_s}{E(J-K_s)}$=0.69]{cardelli89} 
and by \citet[][$\frac{AK_s}{E(J-K_s)}$=0.49]{majaess2016}, 
to derive the DM$_0$. We found, respectively, DM$_0$=13.571$\pm$0.073$\pm$0.049 mag
and DM$_0$=13.718$\pm$0.075$\pm$0.052 mag. 
The use of different reddening laws affects the estimate 
of the true distance modulus at the level of 
$\pm$1$\sigma$, but in opposite directions. Therefore, we will not take
into account the DMs derived from the PW and simply adopt the average value
from the PLs as our final estimate of DM$_0$.

We note that the overall estimate agrees within 1$\sigma$ with 
previous estimates based on RRLs, but using the same 
photometry \citep{braga16,braga2018}.

\section{Summary and final remarks}\label{section_conclusion}
%_______________________________________________________

We have adopted Optical ($UBVRI$) and NIR ($JHK_s$) PSF
photometry of \wcen~\citep{braga16,braga2018} 
and derived the pulsation properties 
(periods, mean magnitudes, light amplitudes and light-curve
Fourier coefficients) of its seven T2Cs. We have discussed in detail the 
transition between RRLs and T2Cs, by using also 
data from the OGLE-IV survey and adopting several
diagnostics (CMD, Bailey diagram, Fourier 
coefficients and light-curve template). We found that the 
period threshold at 1 day---commonly adopted
to separate RRLs and T2Cs---is not universal.
This was already evident in the 
literature \citep{sandage1994} but there has been a long-lasting
lack of investigation on this issue. After 25 years, based on
an unprecedented amount of data including the 
OGLE -IV time series \citep{udalski2015} we have obtained
the following results. 

Three main mechanisms severely hamper the 
reliability of the period threshold as a method to separate 
between RRLs and T2Cs.

{\it Evolutionary effects}: RRLs with periods 
longer than 1 day, overlapping with the 
shortest-period T2Cs, are predicted by pulsation models 
for evolved objects approaching helium exhaustion
\citep{marconi15,marconi2018}.\\

{\it Helium abundance}: Although there is no 
direct evidence of helium enhancement of the RRLs
and T2Cs of \wcen, pulsation models predict that 
helium-enhanced RRLs have periods longer than 1 day
\citep{marconi2018}. This means that also helium-enhanced 
RRLs can overlap in period with the shortest-period T2Cs.\\

{\it Period aliasing}: The most severe periodicity alias 
for ground-based observations, when using various types 
of techniques to estimate periodicity (e.g., Lomb-Scargle,
Phase Dispersion Minimization, string length) 
is at 1 day, because of the daily cycle of telescope
activity. The lack of variable stars 
around this period is, at least in part, due to this alias, which 
makes more difficult to detect the correct period. Among the largest
surveys, only OGLE (thanks to its huge number of observations) 
and Gaia \citep{gaia_alldr}, the latter being a space telescope,
easily overcome this limitation.
In the future, LSST is also expected to be less affected by this 
period alias, thanks to the extension of its final time series.\\

To sum up, the period threshold is not universal and, 
when adopted, it brings to an approximate separation.
Based on the diagnostics discussed in Section~\ref{section_rrlt2c} (especially the 
$\phi_{31}$-$\log{P}$ diagram and the residuals from template 
light curves), we propose to re-classify V92 (P=1.346 days)
as a candidate RRab. 
We also found that 15 Bulge variables, previously 
classified as BLHs in the OGLE survey \citep{soszynski2017} are 
more similar to long-period RRab than to other T2Cs. Therefore, 
we suggest to re-classify these variables as candidate RRab stars.

We therefore suggest a more rigorous---although 
less immediate---approach to separate 
RRLs and T2Cs, based on the evolutionary status of
the star. We consider as RRLs all the 
pulsating stars of the RRL and Cepheid IS, that are 
in their core helium-burning stage, regardless of their period.
Only those stars that have exhausted helium in their cores should
be classified as T2Cs.

Unfortunately, it is not easy to provide a solid 
diagnostic to follow this evolutionary criterion, for several reasons.
$i)$---The ZAHB and helium-exhaustion tracks in the CMD do depend 
on metallicity and helium abundance. Moreover, each 
stellar evolutionary code generates slightly different tracks, 
depending on model assumptions (e.g. convection efficiency).
On top of this, a proper comparison with empirical data would require 
a supplementary spectroscopic investigation to estimate 
the [Fe/H] abundance of the T2Cs.
$ii)$---Although the $\phi_{31}$-$\log{P}$ diagram is very informative, 
we have checked that the morphological subclasses of BLHs 
(the so-called AHB1, AHB2 and AHB3, where AHB1 are classically
associated to RRab) are not well-separated in this
diagram. Moreover, one needs at least some tens of phase points to derive 
a reliable estimate of the Fourier coefficients, which is not always
the case for extensive surveys where variability is not among the main
science cases.

We have studied the properties of the 
PL relations of RRLs and T2Cs. 
We found empirical evidence that RRab and T2Cs 
obey the same $JHK_s$-band PL relations, 
thus confirming the preliminary
working hypothesis by \citet{majaess2010}. 
This has remarkable consequences for distance
estimates and, in turn, the setting of an extragalactic
distance scale anchored only to Population II stars. 
In fact, the most severe limitation to the use 
of RRLs as distance indicators is their faintness, 
despite being ubiquitous and very numerous. 
T2Cs are 1 to 5 mag brighter, meaning that they 
can be detected in both farther and more reddened 
environments. On the other hand, the use of T2Cs is 
hampered by their modest number (at least one order
of magnitude smaller than RRLs).
Therefore, by virtue of the existence of a common PLs,
one could employ RRLs and T2Cs together---as if they were
the same class of variable stars---and overcome their respective
weaknesses as distance indicators.
Despite a more solid calibration --based on more 
objects-- is needed, this assumption opens the path to 
adopt a RRL+T2C calibration of SNIa, leading to an
independent estimate of $H_0$.

Although, at the time of writing, there is no possibility to 
calibrate the RRL+T2C NIR PL relations based on a large
sample of both types of variables, instrumentation 
coming up in the near future will provide this opportunity. First of 
all, WFIRST and JWST \citep{jwst_main} will provide NIR photometry of, respectively,
wide/shallow and narrow/deep areas. Also, the next
Gaia data releases will provide not only more accurate parallaxes
for a geometrical calibration, but also more extended time series. 
On the other hand, LSST \citep{lsst_main} will provide an unprecedented 
wealth of time series in six passbands ($ugrizy$). These 
will be crucial to establish more quantitative criteria
to separate RRLs from T2Cs, by using both the quoted 
diagnostics and, eventually, the colour-colour curves
\citep{diethelm1983} for which an optimal coverage of the 
light curves in at least three passbands is required.

\section*{Acknowledgements}
We thank the anonymous refere for his/her valuable suggestions, 
which helped to improve the content and shape of the paper.
V.F.B. acknowledges the financial support of the Istituto Nazionale di 
Astrofisica (INAF), Osservatorio Astronomico di Roma, and 
Agenzia Spaziale Italiana (ASI) under contract to INAF: 
ASI 2014-049- R.0 dedicated to SSDC.   
G.F. has been supported by the Futuro in Ricerca 2013 (grant RBFR13J716). 

\begin{appendix}

\section{Identification}\label{chapt_rrind_omega}

We did not found any new T2C within our new data, 
so we will consider only the seven ones in the 
Clement catalogue \citep{Clement01} and 
classified as Population II Cepheids by \citet{kaluzny04}. 
We will discuss the transition between
long-period RRLs and BLHs in Section~\ref{section_rrlt2c}.

% %grep -v kal V23/V23_1_allnew_kal9704.fas | awk '{sum+=$1} END {if (NR>0) print sum/NR}'
\begin{table*}
\scriptsize
\caption{Astrometric properties and classification of the T2Cs in \wcen.}
\label{tbl:coord}
\centering
\begin{tabular}{l l l cc cc cc}
\hline
\hline
ID & type & AHB class & RA & Dec & RA$_{Gaia}$ & Dec$_{Gaia}$  & $\mu^{RA}_{Gaia}$ & $\mu^{Dec}_{Gaia}$ \\ 
& & & deg & deg& deg & deg & mas/yr & mas/yr \\
\hline
\wcen\tablefootmark{a} &         &\ldots& 201.694625 & --47.483306 & 201.697    & --47.480    & --3.234$\pm$ 0.039 &  --6.719$\pm$  0.039  \\ 
     V1 & RVT                             & N/A  & 201.521542 & --47.395194 & 201.521533 & --47.395181 & --3.18 $\pm$  0.12 &  --7.79 $\pm$  0.19  \\ 
    V29 & WV                              & N/A  & 201.613458 & --47.479889 & 201.613428 & --47.479890 &  \ldots            &   \ldots             \\ 
    V43 & BLH                             & AHB1 & 201.643792 & --47.449389 & 201.644919 & --47.449480 &  \ldots            &   \ldots             \\
    V48 & BLH                             & N/A  & 201.657542 & --47.507083 & 201.657495 & --47.507090 & --3.57 $\pm$  0.12 &  --7.11 $\pm$  0.25  \\
    V60 & BLH                             & AHB3 & 201.648708 & --47.546806 & 201.648682 & --47.546805 &   \ldots           &   \ldots             \\
    V61 & BLH                             & AHB2 & 201.808208 & --47.458639 & 201.808196 & --47.458634 & --2.48 $\pm$  0.06 &  --7.08 $\pm$  0.10  \\
    V92 & Candidate RRab\tablefootmark{b} & AHB3 & 201.561750 & --47.354139 & 201.561742 & --47.354131 & --2.84 $\pm$  0.07 &  --6.74 $\pm$  0.10  \\\hline
\hline
\end{tabular}
\tablefoot{Col. 1: Name; Col. 2: Variable type;
Col. 3: Light curve morphology type 
(see Section \ref{par:omegacen_lightcurves});
Col. 4: Right ascension (our astrometry); 
Col. 5: Declination (our astrometry); 
Col. 6: Right ascension (Gaia DR2); 
Col. 7: Declination (Gaia DR2); 
Col. 8: Proper motion in Right Ascension (Gaia DR2); 
Col. 9: Proper motion in Declination (Gaia DR2).
The average epoch of our astrometry is 1998.5\\
\tablefoottext{a}{Average for the whole cluster. RA and Dec average from \citet{braga16};
Gaia averages from \citet{vasiliev2019}.}
\tablefoottext{b}{See Section~\ref{section_shortt2c}}.}
\end{table*}

We retrieved all the seven T2Cs in our
images by cross-matching our astrometric solution \citep[with an accuracy
of 0.1\sec, see][]{braga16} with the coordinates from the \citet{Clement01} 
catalogue. We provide updated RA and DEC of the T2Cs of \wcen~in 
Table~\ref{tbl:coord}. We have matched our coordinates with the Gaia DR2 \citep{gaia_dr2}
astrometric catalogue and found that they all match within 0.12\sec, 
except V43. Its {\it Gaia} coordinates are $\sim$2.8\sec away from ours. However,
V43 has no measurements in the $Bp$ and $Rp$ bands, and 
its {\it Gaia} astrometric solution has only 2 parameters (RA,DEC instead
of the more accurate 5-parameter (RA, DEC, $\mu_{RA}$, $\mu_{DEC}$, $\pi$,) solution.
Therefore, we assume that the best coordinates for V43 are ours.

\section{Light curves}\label{par:omegacen_lightcurves}

%FIGURE DELLE LCVS... separate optical e NIR
% temp_190309.pro
\begin{figure}[!htbp]
\centering
% \figurenum{1}
\includegraphics[width=4.3cm]{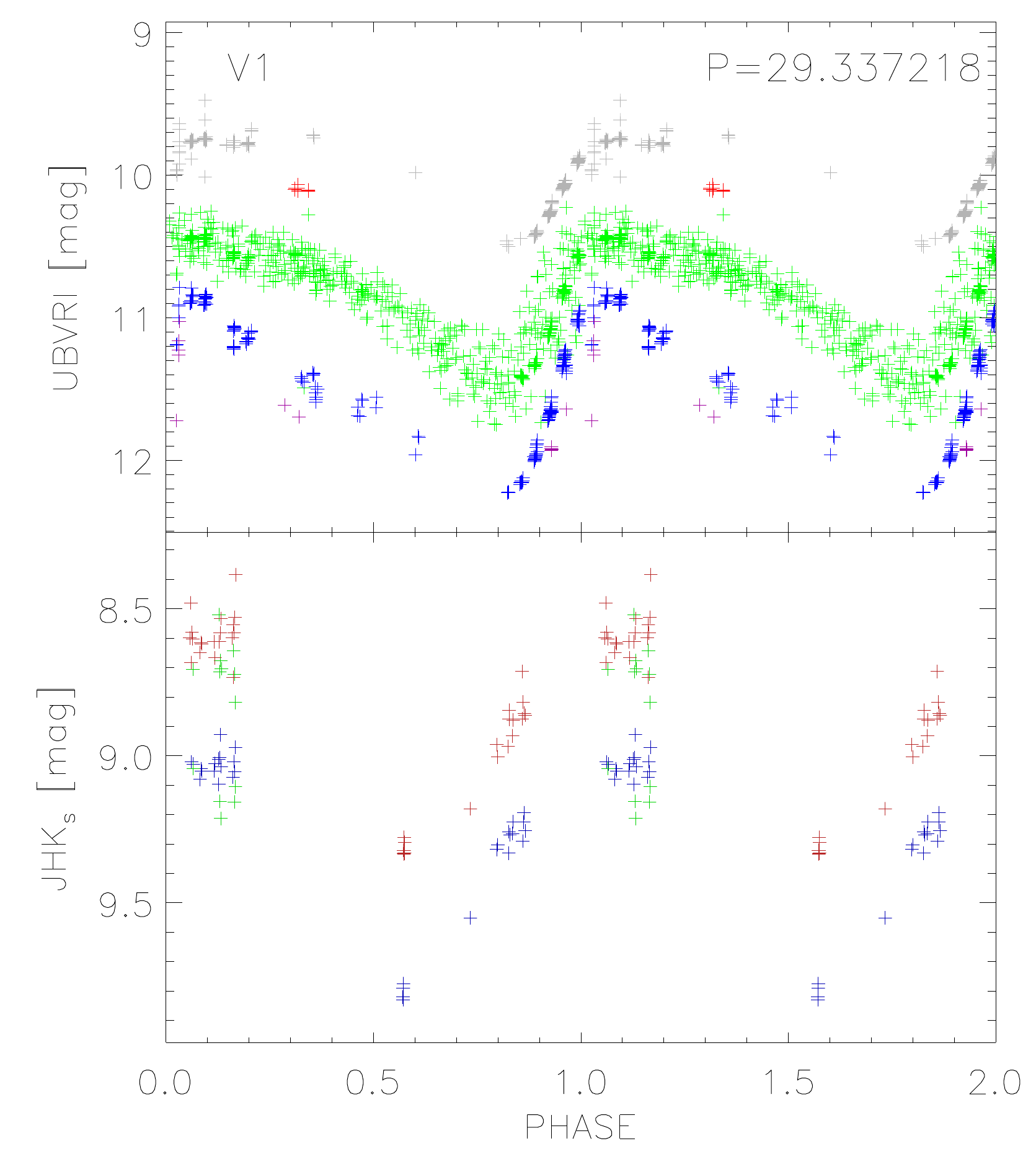}
\includegraphics[width=4.3cm]{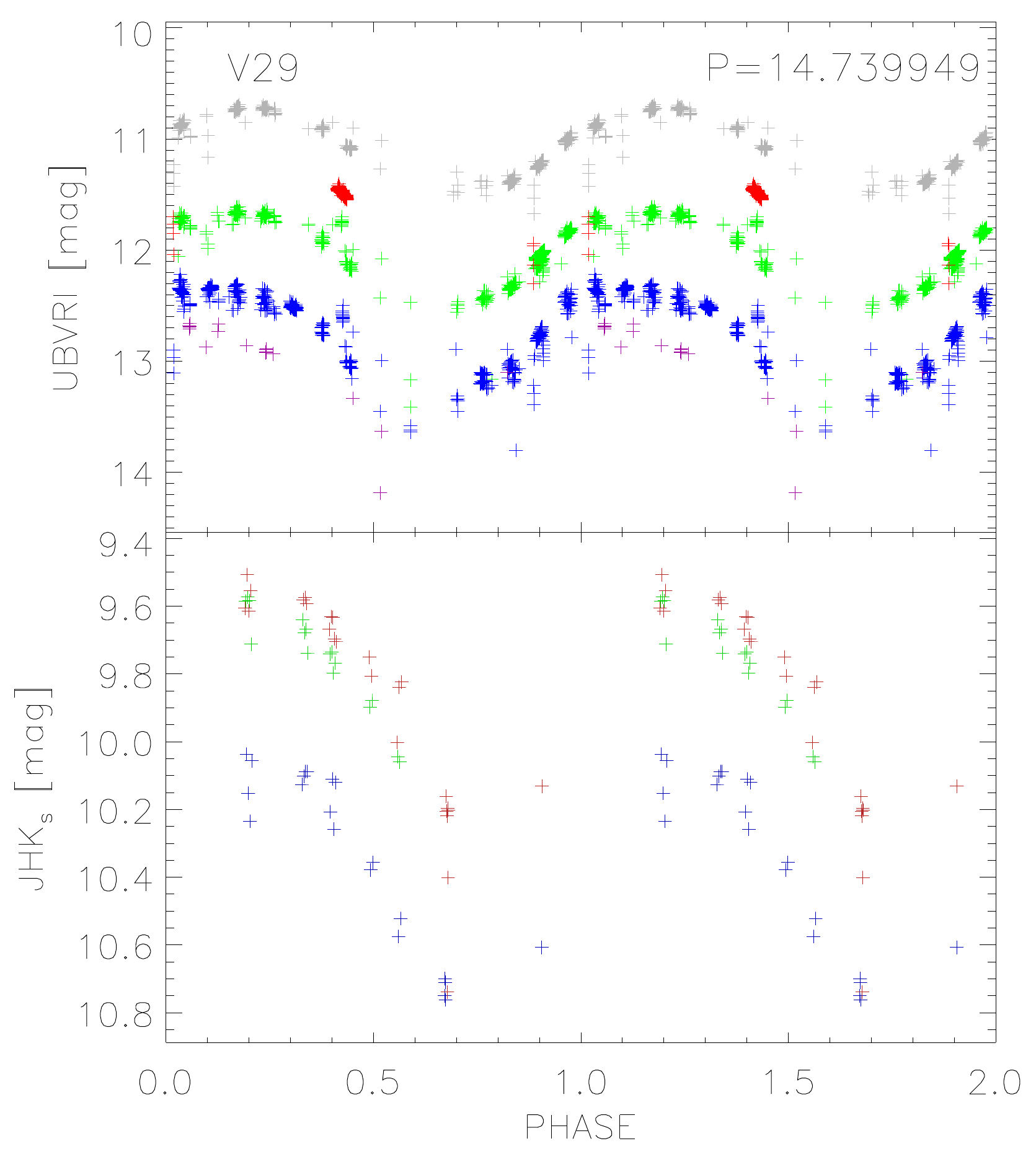}
\includegraphics[width=4.3cm]{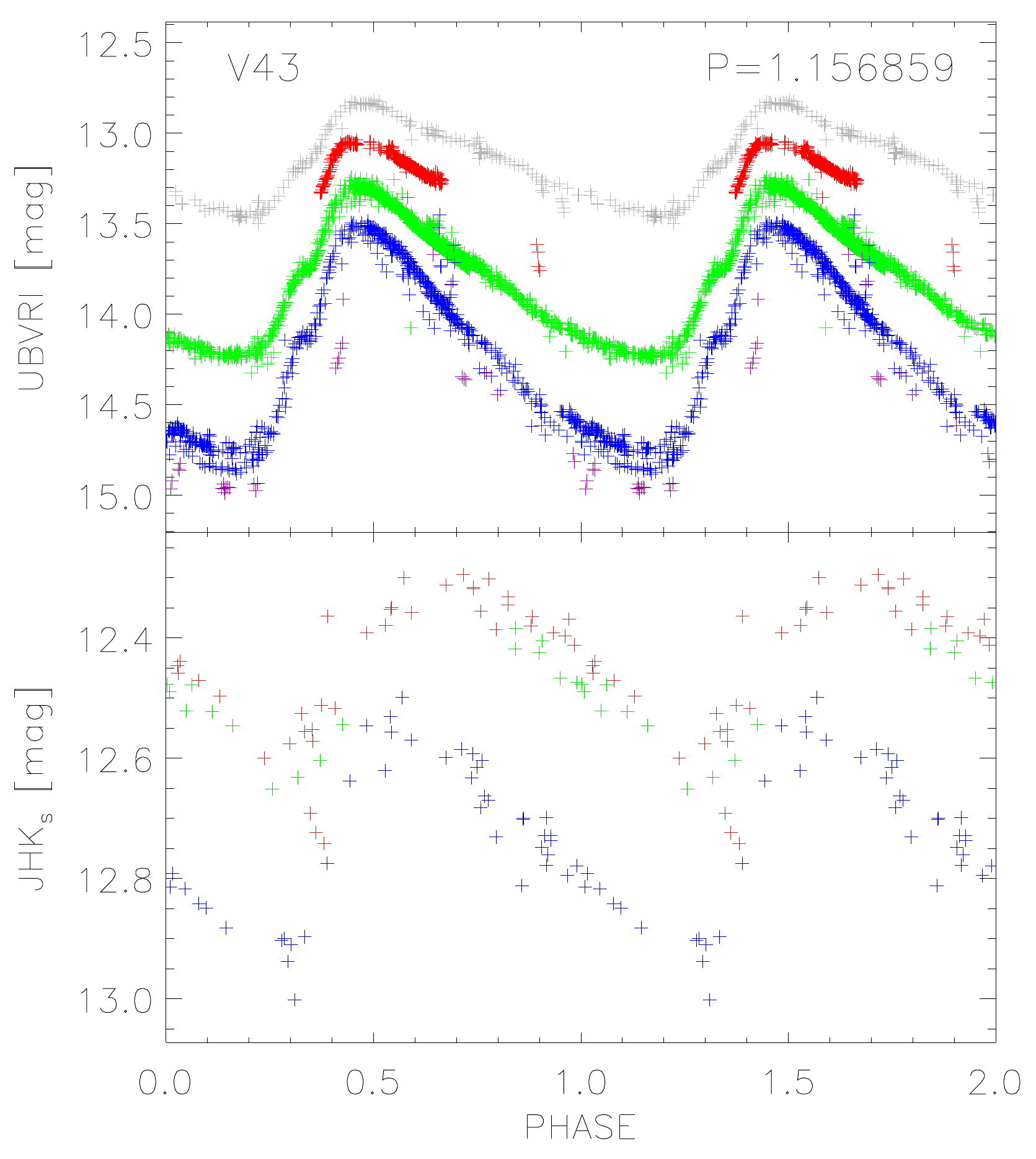}
\includegraphics[width=4.3cm]{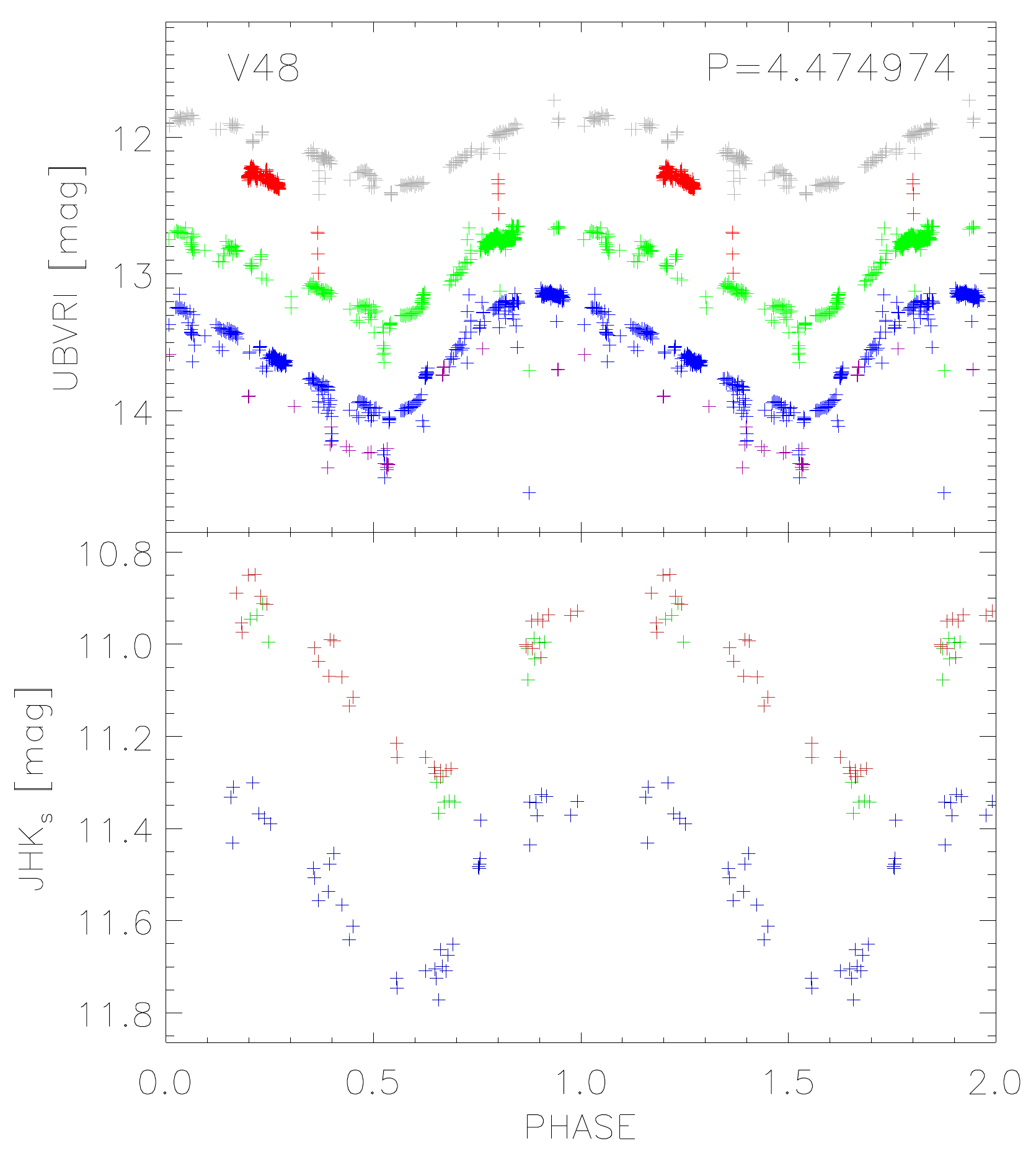}
\includegraphics[width=4.3cm]{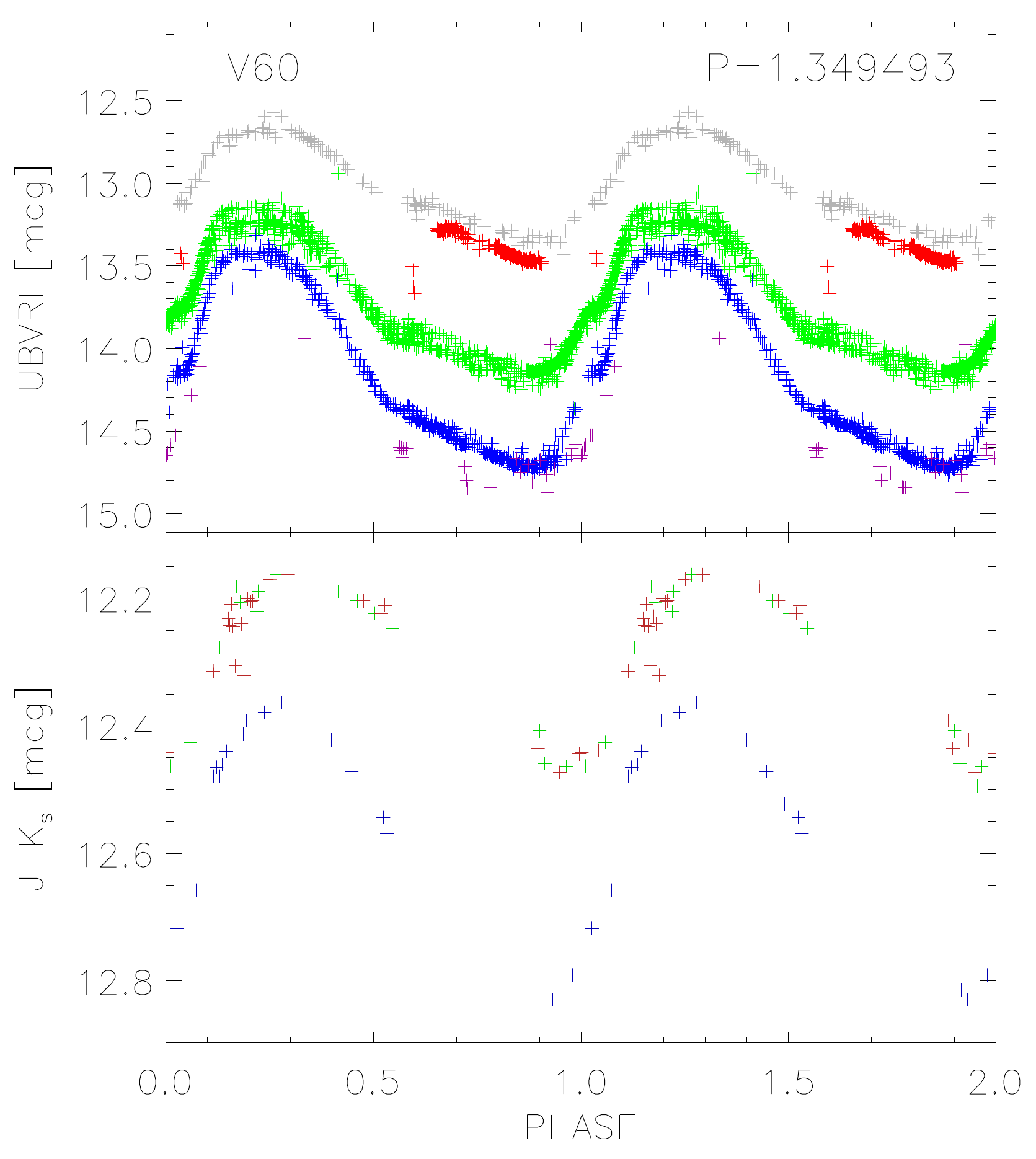}
\includegraphics[width=4.3cm]{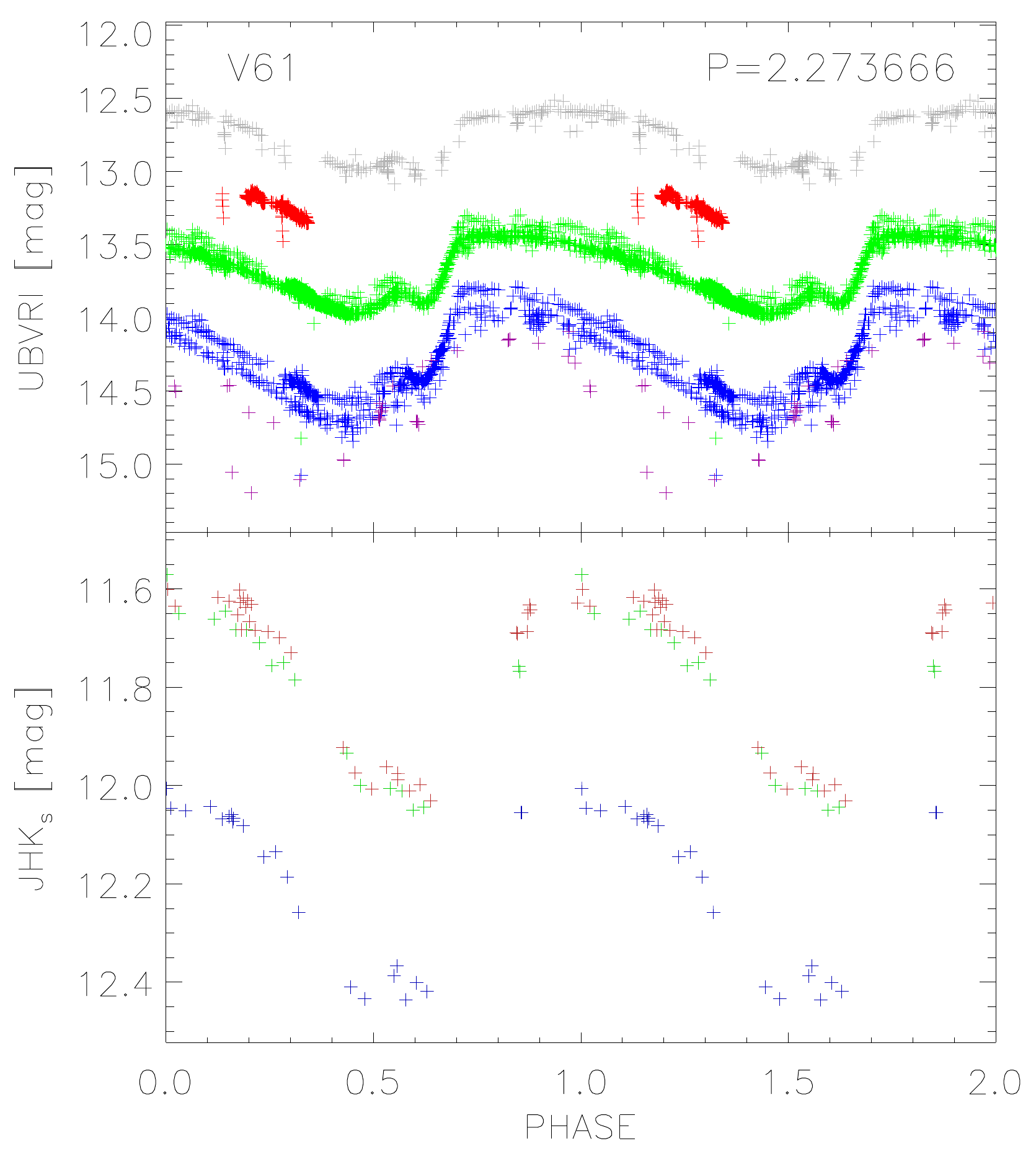}
\includegraphics[width=4.3cm]{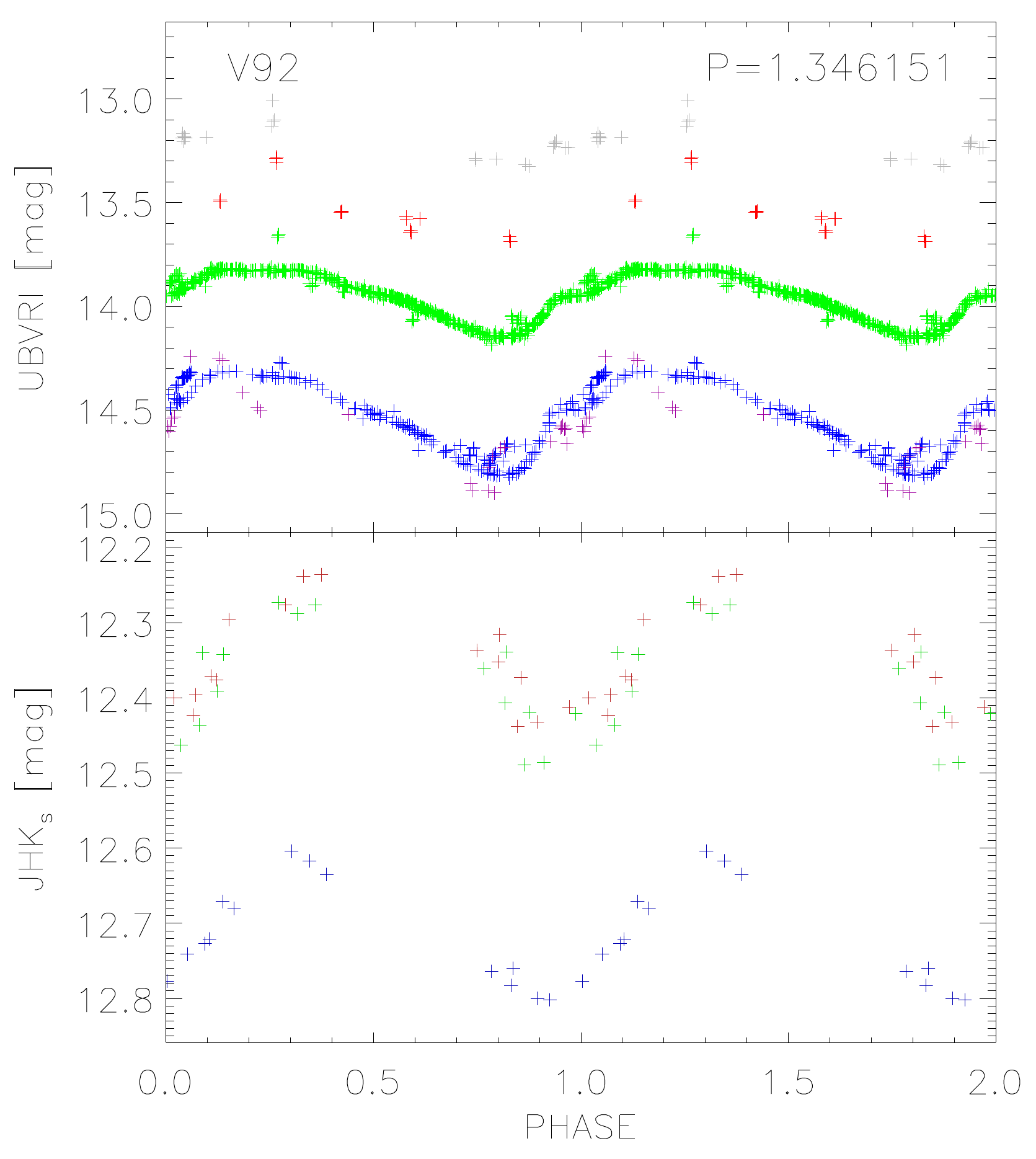}
\vspace*{1.0truecm}
\caption{Top panels: optical light curves of the 
T2Cs of \wcen. Purple: $U$; blue: $B$; green: $V$; red: $R$; 
grey: $I$. Names and periods are labelled at the 
top-left and top-right corners, respectively.
Bottom panels: NIR light curves of the 
T2Cs of \wcen. Blue: $J$; green: $H$; red: $K_s$.}
\label{fig:lcvs}
\end{figure}

For the optical and NIR light curves we have adopted
the photometry by \citet{braga16,braga2018}. 
For V1, we have added photometry from
the ASAS Survey \citep{pojmanski1997}.

\begin{figure}[htbp]
\centering
\includegraphics[height=6cm]{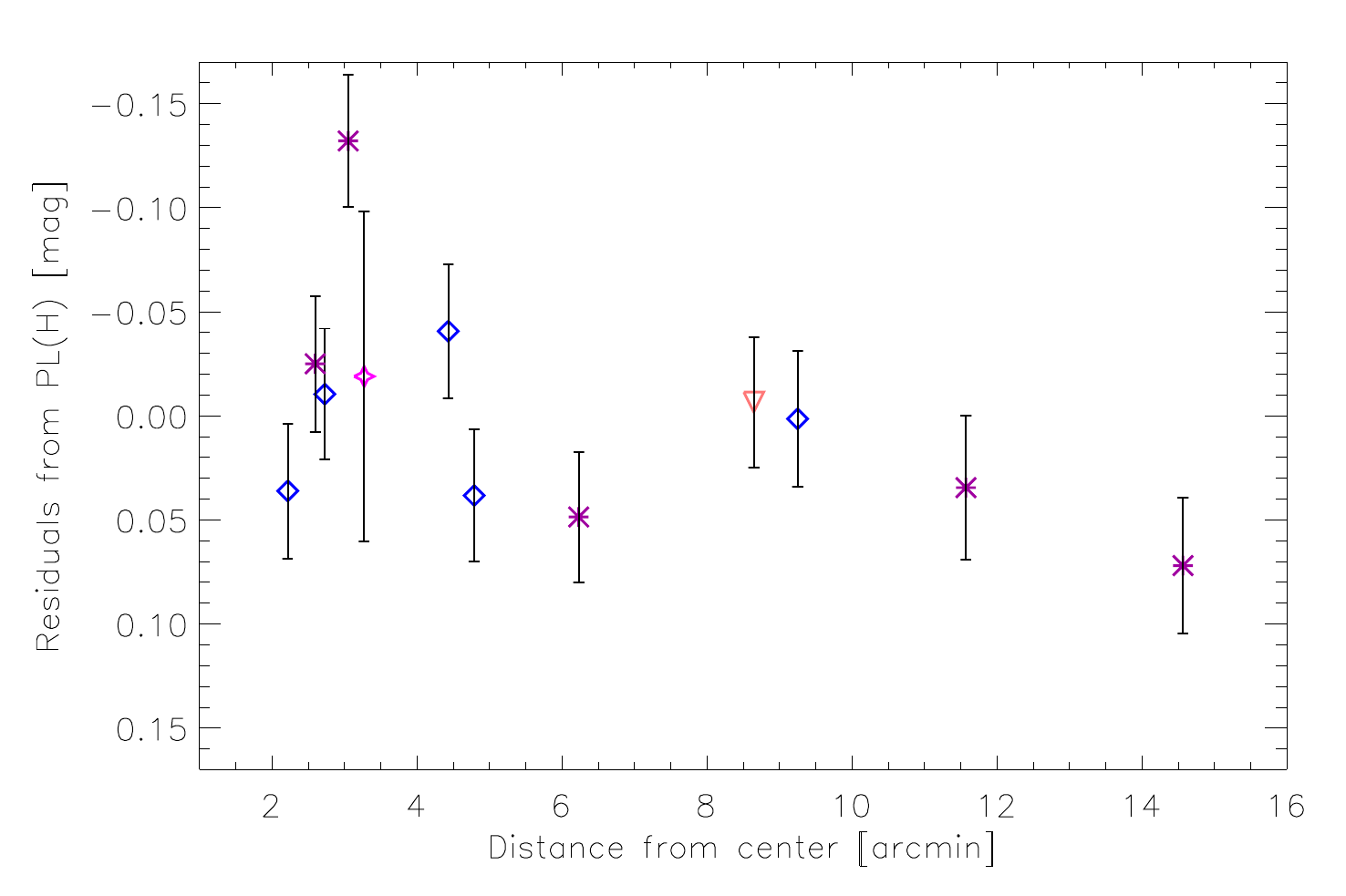}
\caption{Residuals between measured and predicted 
$H$-band magnitude versus angular distance from the center.
Symbols have the same meaning as in Figs.~\ref{fig:basti} 
and~\ref{fig:bailey_field}. The bars represent the squared
sum of photometric error plus the PL intrinsic dispersion.}
\label{fig:blend}
\end{figure}

By following a suggestion of the referee, we have 
quantitatively checked the impact of blending. Fig.~\ref{fig:blend}
displays the residuals of the $H$-band magnitudes of T2Cs and 
long-period RRLs respect to the empirical PL($H$) derived in 
Section~\ref{chapt_t2c_pl}, plotted versus the angular
distance from the center of the cluster. The small sample
and wide dispersion of the residuals do not allow to give 
firm conclusions. However, there is no clear trend suggesting that, 
within the errors, blending does not affect our estimates.

Figure~\ref{fig:lcvs} displays the optical and NIR 
phased light curves of the T2Cs. Their periods are displayed
in Table~\ref{tbl:magnitudes} and were derived performing the multiband 
approach described in \citet{stetson14a}, using both optical and 
NIR data. We did not find any significant difference between our periods
and those in the \citet{Clement01} catalogue, in fact the relative 
offsets are smaller than 5$\cdot 10^{-4}$ for all the T2Cs. 

V1 does not show clear signs of alternating
deep and shallow minima, typical of field---presumably 
intermediate-mass---RVTs. Moreover, it does follow
the same PL and PW relations as BLHs and WVs 
(see Section~\ref{chapt_comparison_cluster}), which are the 
common features of GGC RVTs. According to 
\citet{gonzalez1994b}, it is also 
metal-poor ([Fe/H]=--1.77 dex) and highly enhanced
in $\alpha$ elements ([$\alpha$/Fe] $>$0.4 dex). Therefore, 
we will consider V1 as a bona-fide (low-mass and old) T2C.

V29 displays an almost sinusoidal light curve, typical of 
WVs with similar periods. It shows no signs of 
over-luminosity, therefore it should be classified as a 
bona-fide WV.

BLHs, on the other hand, display varied light-curve morphologies, 
also at similar periods (see Fig.\ref{fig:lcvs}).
This is a known issue: \citet{diethelm1990} and
\citet{sandage1994} 
defined three different classes of T2Cs with periods
shorter than 3 days, namely, AHB1, AHB2 and AHB3 (See 
Appendix~\ref{appendix2}).
Due to the wide magnitude range covered by T2Cs, and 
in turn, the broad range of evolutionary phases they 
experience, the classification of their light curve morphology 
and their correlation with physical parameters (stellar 
mass, chemical composition) is difficult. 
We inspected the light curve morphologies of the BLHs in \wcen.
The results are the following, as  
summarized in Table~\ref{tbl:coord}:

V43)---Its light curve displays a steep rising branch
and a shallower decreasing branch, typical of AHB1 stars.

V48)---Its light curve is almost 
sinusoidal and does not look like any of the quoted morphological 
types. This is not unexpected, because its period 
is $\sim$4.5 days, which is outside of 
the period range of the AHBs. It is the brightest BLH in \wcen~and the 
only one for which elemental abundances were derived \citep{gonzalez1994b}.
Its low metallicity ([Fe/H]=--1.66) and the enhancement in $\alpha$ elements
suggest that it is of the spectroscopic class UY Eri, recently 
proposed by \citep{kovtyukh2018b}.

V60)---The light curve displays a flat maximum, typical of AHB3 stars.

V61)---The light curve displays the typical bump of AHB2 class. 

V92)---The light curve displays a flat maximum, typical of AHB3 stars.

\begin{table*}
\scriptsize
\caption{Optical mean magnitudes, amplitudes and periods of the T2Cs in \wcen.}
\label{tbl:magnitudes}
\centering
\begin{tabular}{l@{\hspace{1.8mm}} c@{\hspace{1.8mm}}c@{\hspace{1.8mm}}c@{\hspace{1.8mm}}c@{\hspace{1.8mm}}c@{\hspace{1.8mm}}c@{\hspace{1.8mm}} 
c@{\hspace{1.8mm}}c@{\hspace{1.8mm}}c@{\hspace{1.8mm}}c@{\hspace{1.8mm}}c@{\hspace{1.8mm}}}
\hline
\hline
ID & Period & $U$& $B$& $V$& $R$& $I$& Amp($U$)& Amp($B$)& Amp($V$)& Amp($R$)& Amp($I$) \\
  & days & mag & mag & mag & mag & mag & mag & mag & mag & mag & mag \\
\hline
     V1  &  29.337218 &  11.923$\pm$0.484 &  11.488$\pm$0.011 &  10.829$\pm$0.029 &  10.102$\pm$0.014 &  10.058$\pm$0.012 &     \ldots          &   1.355$\pm$0.062 &   1.025$\pm$0.033 &    \ldots       & 0.767 $\pm$ 0.071 \\ 
    V29  &  14.739949 &  12.867$\pm$0.165 &  12.776$\pm$0.014 &  12.015$\pm$0.007 &  11.507$\pm$0.034 &  11.049$\pm$0.018 &     \ldots          &   1.226$\pm$0.143 &   0.968$\pm$0.013 &    \ldots       & 0.795 $\pm$ 0.024 \\ 
    V43  &  1.1568588 &  14.467$\pm$0.016 &  14.139$\pm$0.024 &  13.759$\pm$0.008 &  13.175$\pm$0.071 &  13.149$\pm$0.031 &   1.084$\pm$0.056   &   1.235$\pm$0.038 &   0.926$\pm$0.014 &    \ldots       & 0.632 $\pm$ 0.036 \\ 
    V48  &  4.4749736 &  13.845$\pm$0.014 &  13.528$\pm$0.017 &  12.924$\pm$0.008 &  12.317$\pm$0.037 &  12.092$\pm$0.012 &   0.855$\pm$0.130   &   0.853$\pm$0.032 &   0.667$\pm$0.011 &    \ldots       & 0.611 $\pm$ 0.055 \\ 
    V60  &  1.3494930 &  14.353$\pm$0.025 &  14.028$\pm$0.017 &  13.624$\pm$0.007 &  13.420$\pm$0.078 &  13.001$\pm$0.071 &   0.938$\pm$0.150   &   1.298$\pm$0.035 &   1.006$\pm$0.014 &    \ldots       & 0.665 $\pm$ 0.081 \\ 
    V61  &  2.2736663 &  14.487$\pm$0.017 &  14.293$\pm$0.004 &  13.661$\pm$0.003 &  13.254$\pm$0.064 &  12.821$\pm$0.013 &   0.890$\pm$0.131   &   0.773$\pm$0.027 &   0.548$\pm$0.009 &    \ldots       & 0.365 $\pm$ 0.029 \\ 
    V92  &  1.3461514 &  14.590$\pm$0.016 &  14.480$\pm$0.006 &  13.946$\pm$0.003 &  13.567$\pm$0.037 &  13.199$\pm$0.011 &   0.482$\pm$0.056   &   0.438$\pm$0.027 &   0.323$\pm$0.008 & 0.207$\pm$0.057 & 0.184 $\pm$ 0.028 \\ 
\hline
\end{tabular}
\end{table*}

\begin{table*}
\scriptsize
\caption{NIR photometric properties of the T2Cs in \wcen.}
\label{tbl:nirmagnitudes}
\centering
\begin{tabular}{l@{\hspace{1.8mm}}c@{\hspace{1.8mm}}c@{\hspace{1.8mm}}c@{\hspace{1.8mm}}c@{\hspace{1.8mm}} 
c@{\hspace{1.8mm}}c@{\hspace{1.8mm}}c@{\hspace{1.8mm}}}
\hline
\hline
ID & $J$& $H$& $K_s$& Amp($J$)& Amp($H$)& Amp($K_s$) \\
 & mag & mag & mag & mag & mag & mag \\
\hline
     V1  &  9.334$\pm$0.022 &   9.008$\pm$0.005 &   8.879$\pm$0.023 & 0.788$\pm$0.083 &       \ldots    &   0.743$\pm$0.064  \\ 
    V29  & 10.379$\pm$0.013 &   9.736$\pm$0.073 &   9.854$\pm$0.026 & 0.738$\pm$0.054 &       \ldots    &   0.757$\pm$0.078  \\ 
    V43  & 12.730$\pm$0.013 &  12.492$\pm$0.006 &  12.426$\pm$0.013 & 0.393$\pm$0.040 & 0.252$\pm$0.035 &   0.274$\pm$0.029  \\ 
    V48  & 11.470$\pm$0.013 &  11.078$\pm$0.009 &  11.034$\pm$0.011 & 0.433$\pm$0.037 & 0.438$\pm$0.053 &   0.408$\pm$0.038  \\ 
    V60  & 12.584$\pm$0.005 &  12.295$\pm$0.008 &  12.281$\pm$0.008 & 0.453$\pm$0.053 & 0.301$\pm$0.036 &   0.279$\pm$0.027  \\ 
    V61  & 12.190$\pm$0.007 &  11.811$\pm$0.007 &  11.771$\pm$0.008 & 0.411$\pm$0.038 & 0.425$\pm$0.048 &   0.397$\pm$0.034  \\ 
    V92  & 12.700$\pm$0.004 &  12.340$\pm$0.010 &  12.313$\pm$0.008 & 0.189$\pm$0.025 & 0.210$\pm$0.031 &   0.187$\pm$0.027  \\ 
\hline
\end{tabular}
\end{table*}

Our morphological classification 
matches that provided by \citet{sandage1994}
except for V60 and V92: they were classified, respectively, as  
borderline AHB1 and AC (Anomalous Cepheid). 
We rule out the possibility that V92 is an AC because, based on its 
proper motion (see Table~\ref{tbl:coord}), it is a cluster 
member (the proper motion of \wcen~is 
$\mu^{RA}_{Gaia}$=--3.24$\pm$0.01 mas/yr; 
$\mu^{Dec}_{Gaia}$=--6.73$\pm$0.01 mas/yr,
\citealt{baumgardt2019}) but does not follow the PL relation of 
cluster-member ACs, while it does follow the PL relation of 
cluster-member BLHs (see Section~\ref{chapt_t2c_pl}).

%_______________________________________________________________________________
% \subsection{CMD Opt-nir}\label{chapt_cmd}

% figure generate con cmd_bvi2
% cp ../cmd_bvi_var_final_t2c-eps-converted-to.pdf .
% cp ../cmd_kbk_var_lco_final_t2c-eps-converted-to.pdf .

\begin{figure}[htbp]
\centering
\includegraphics[trim={1.2cm 0 0 0},clip,height=8cm]{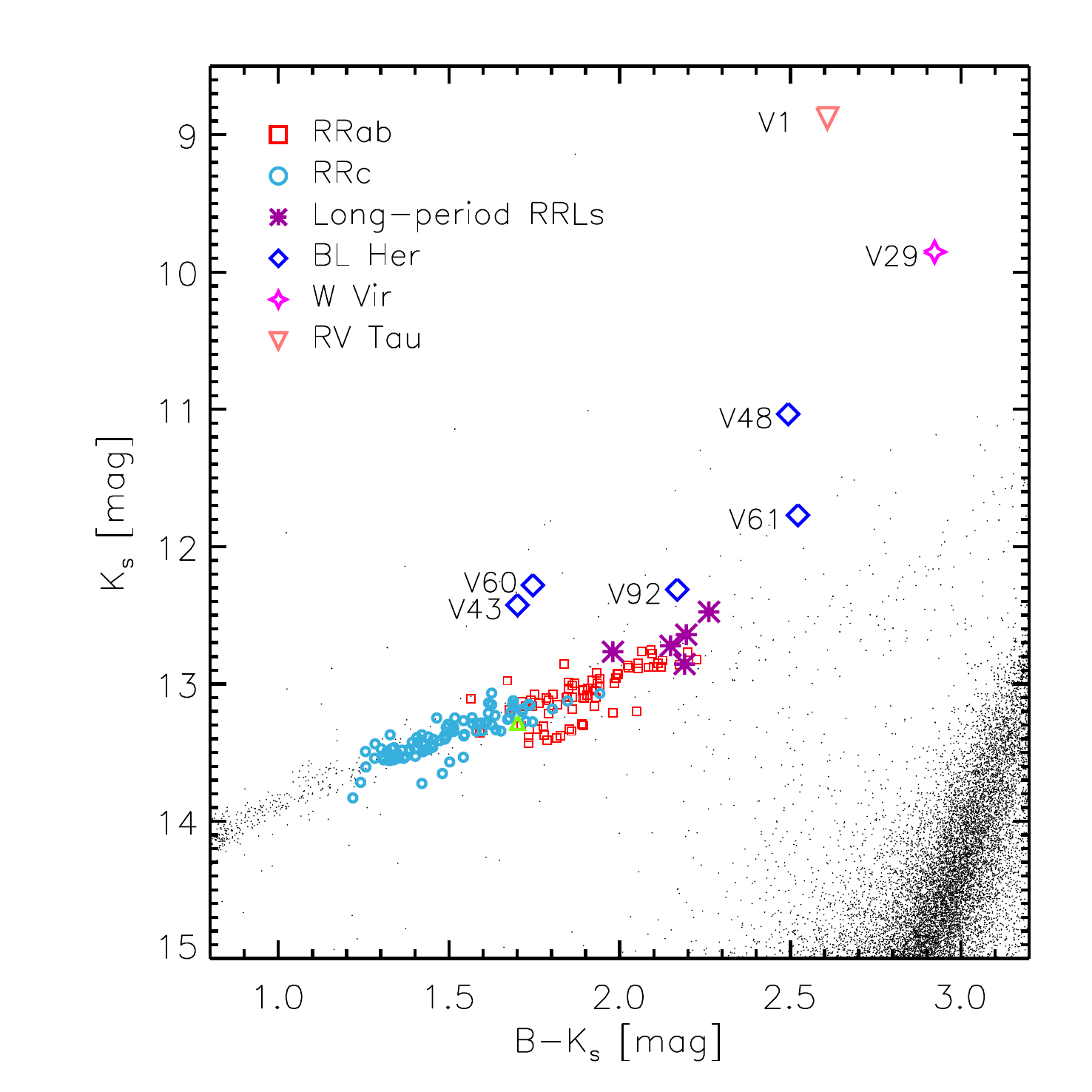}
\caption{
Optical-NIR($K_s$,$B-K_s$) CMD 
of \wcen~(close-up on RRLs and T2Cs).
Symbols have the same meaning as in Figs.~\ref{fig:basti} 
and~\ref{fig:bailey_field}.}
\label{fig_cmd}
\end{figure}

\begin{figure}[htbp]
\centering
\includegraphics[width=8cm]{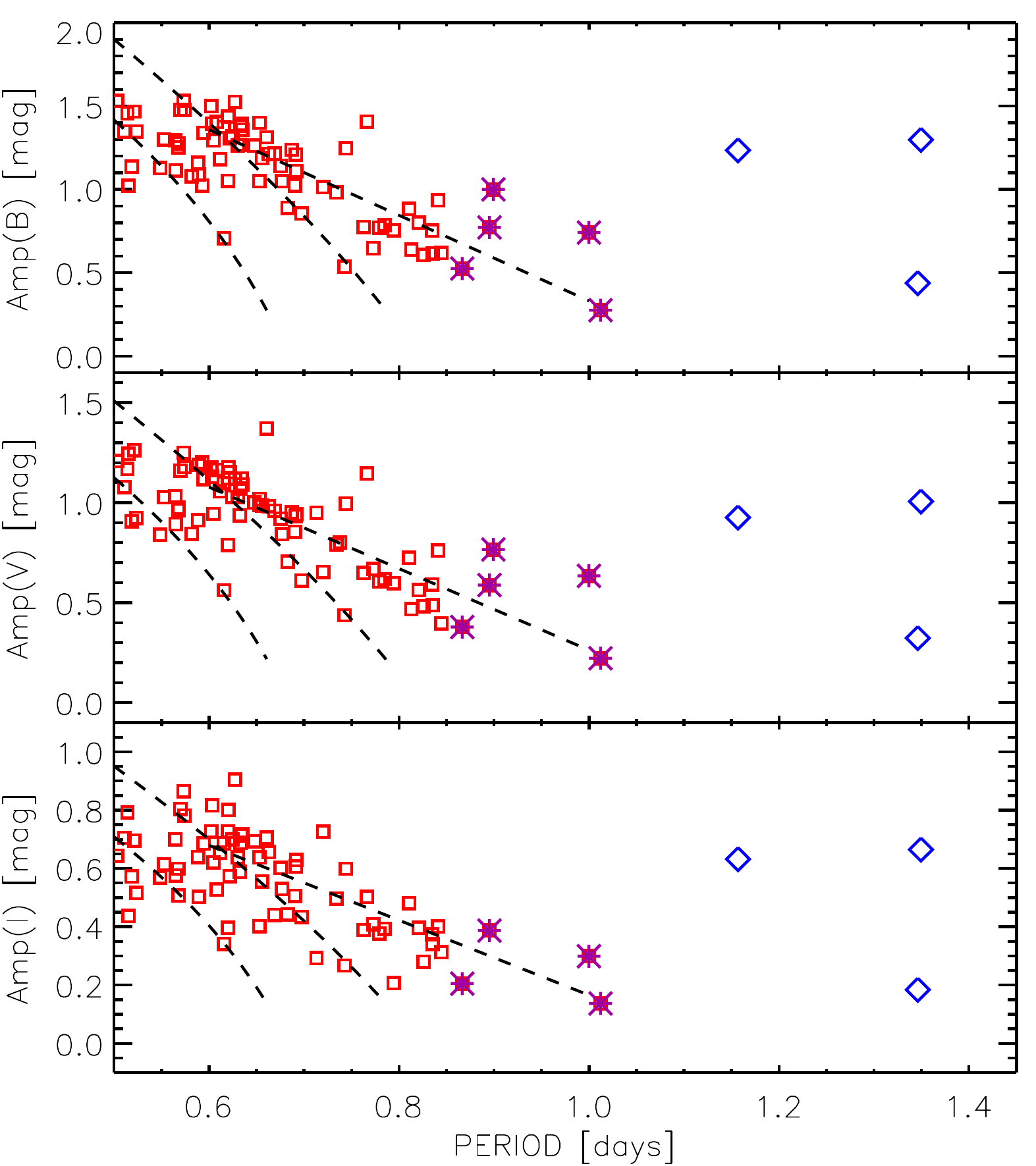}
\caption{Bailey diagram of RRab and T2C stars in \wcen. The 
symbols are the same as in Fig.~\ref{fig_cmd}.
Dashed lines represent the Oosterhoff 0 (Oo0, see 
section~\ref{section_longrrl}), Oosterhoff I (OoI) and 
Oosterhoff II (OoII) loci of RRab stars. The Oo0 linear
relation was derived by us in \ref{section_longrrl}. OoI 
and OoII were adopted from \citet{fabrizio2019}. We rescaled
the relations by using the amplitude ratios provided
by \citet{braga16}.}
\label{fig:bailey}
\end{figure}

%V52a: neighbor at 0.5''. Blended in danish datasets. It is possible that even the {\it other} dataset is not properly deblended
% V68: brightest RRc. magnitudes seem to be accurate. no bright neighbor (try a larger radius).
% V99: P=0.766, good light curve sampling, no bright neighbors.

\section{Above Horizontal Branch variables subclasses}\label{appendix2}

\begin{figure}[htbp]
\centering
\label{fig:ahb}
\includegraphics[width=8cm]{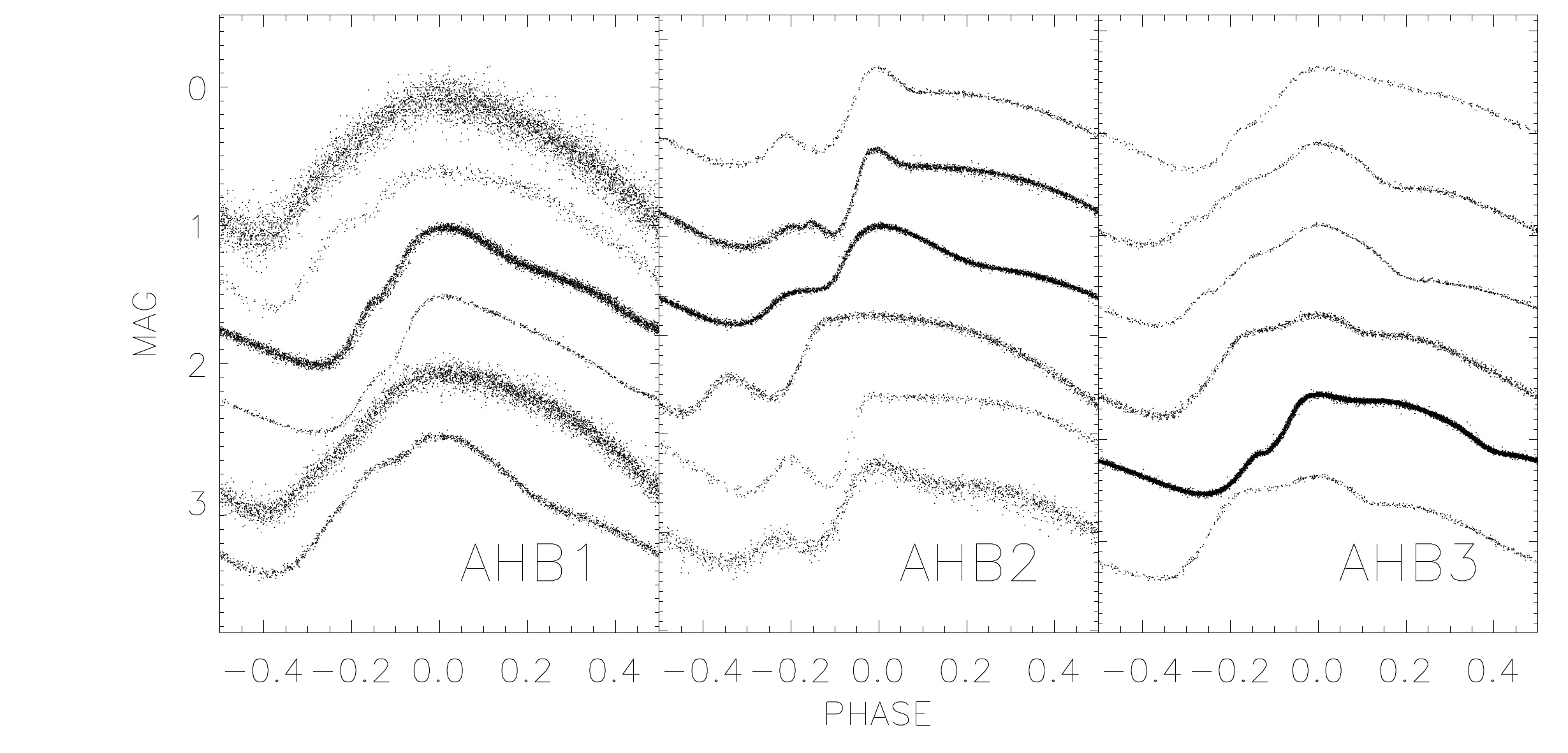}
\caption{Sample normalized $I$-band 
light curves of Bulge AHB variables from OGLE-IV
\citep{soszynski2017}.}
\end{figure}

As \citet{diethelm1983} pointed out, variable AHB stars come
in three different flavors, which were later labelled as 
AHB1, AHB2 and AHB3 by \citet{sandage1994}. 
Their classification is based on the morphology of
their optical light curves (see Fig.~\ref{fig:ahb}).

% figura generata con temp_200430.pro

AHB1 display saw-tooth RRab-like light curves (in fact, 
they were associated to RRLs by \citet{diethelm1983}; 
AHB2 display a prominent secondary peak before the rising branch, and 
AHB3 stars display either a bump on the decreasing branch, or
a plateau which covers $\sim$20\% 
of the pulsation cycle around the maximum light.
Note that the prototype BL Her itself is a 
AHB3 star, with a solar-like metallicity, higher than the
majority of T2Cs \citep{caldwell1978,maas2007}.
We have checked that AHB1 and AHB3 variables are not 
well separated in the Bailey and the 
$\phi_{31}$ vs $\log{P}$ diagrams, while AHB2 are 
more easily detectable.
%However, this is beyond the 
%aim of this paper.

\section{Light curve template of long-period RRLs}\label{sec:append}

\begin{figure}[htbp]
\centering
\label{fig:template}
\includegraphics[width=8cm]{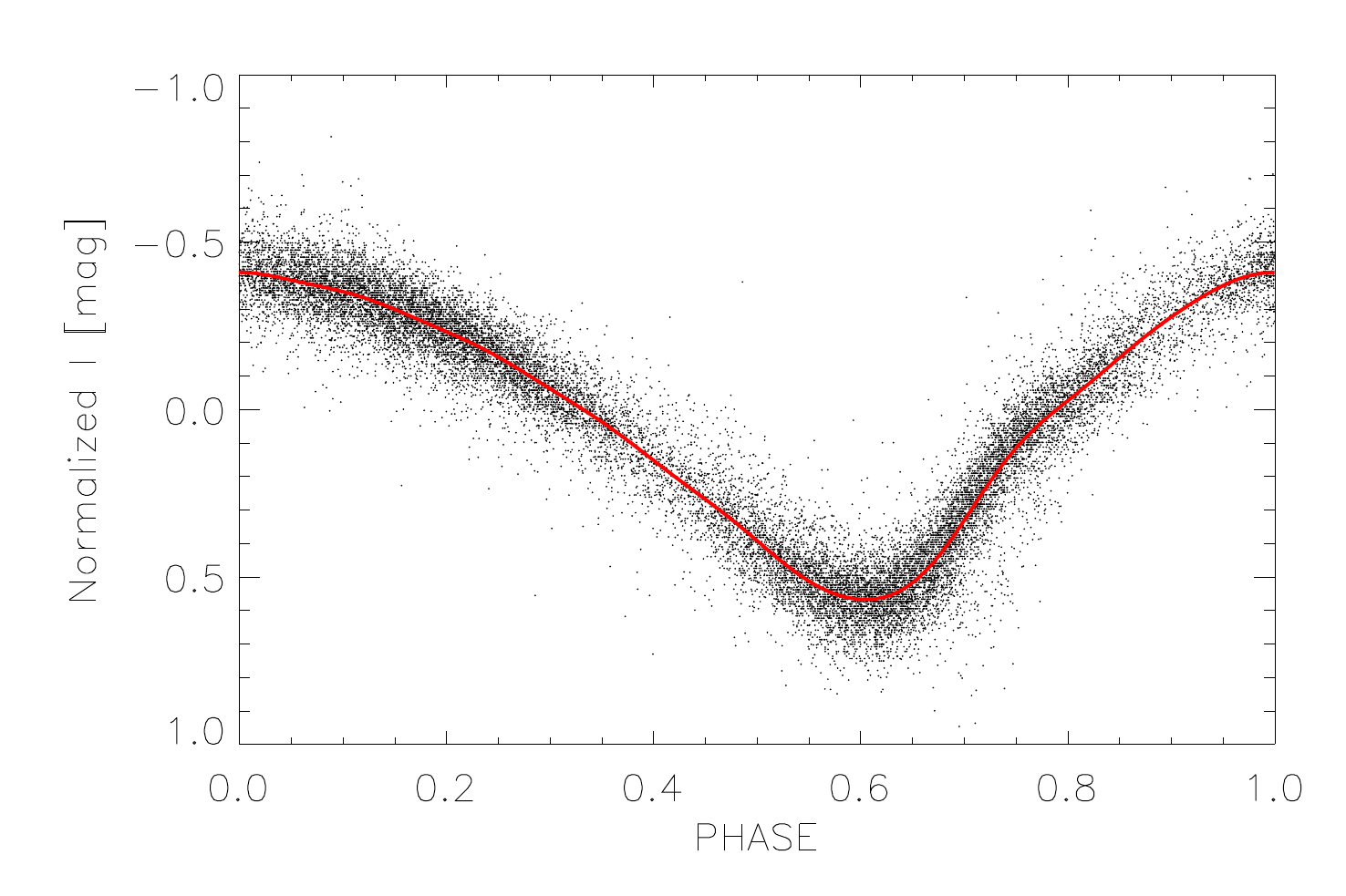}
\caption{Cumulated, normalized light curve of long-period RRLs 
and the Fourier fit adopted as template light curve (red line).}
\end{figure}

To build the $I$-band light curve template of long-period RRLs, 
we have selected 16 RRLs with periods between 0.97 and 1.00 days, 
that is, a sample that has no intersection with either the 
``candidate RRLs'' or the ``Short-period T2Cs'' groups.
We normalized and co-phased their $I$-band light curves by adopting
the $I$-band amplitude and epoch of maximum provided by the OGLE 
collaboration. Their cumulated, normalized light curve was fitted 
with a 8$^{th}$-order Fourier series (see Fig.\ref{fig:template}).

% figura generata con temp200414
% cp /media/vittorio/Volume/Surveys/OGLE/OGLE4RR/BUL/pseudotemplate_ge097-eps-converted-to.pdf .

The coefficients are shown in Table~\ref{tbl:template}.
We adopt this fit as the light-curve template for long-period RRLs.

% \[F(\phi) = A_0 + \Sigma_i A_i \cos{(2\pi i \phi - \phi_i)} \tag{1} \]

\begin{table}
\footnotesize
\caption{Coefficients of the light-curve template.}
\label{tbl:template}
\centering
\begin{tabular}{l r|l r}
\hline
\hline
A$_0$ &  0.00915  &          &  \ldots   \\
A$_1$ &  0.44528  & $\phi_1$ &  2.74455  \\
A$_2$ &  0.08866  & $\phi_2$ &  4.74558  \\
A$_3$ &  0.02648  & $\phi_3$ &  1.19303  \\
A$_4$ &  0.02067  & $\phi_4$ &  3.33520  \\
A$_5$ &  0.00940  & $\phi_5$ &  4.90126  \\
A$_6$ &  0.00279  & $\phi_6$ &  0.06851  \\
A$_7$ &  0.00221  & $\phi_7$ &  2.19867  \\
A$_8$ &  0.00346  & $\phi_8$ &  3.77858  \\
\hline
\end{tabular}
\tablefoot{$F(\phi) = A_0 + \Sigma_i A_i \cos{(2\pi i \phi - \phi_i)}$}
\end{table}

\end{appendix}

\bibliographystyle{aa}
\bibliography{../../../../../Latex/ms}

\end{document}